\documentclass[12pt]{oxarticle}
\pdfoutput=1

\usepackage{graphicx, float, array, xspace, amscd, amsmath, amsthm, amssymb, latexsym, relsize, bbm}
\usepackage[left=1.7in, right=0.3in, top=0.8in, bottom=1.2in]{geometry}
\usepackage[bbgreekl]{mathbbol}
\usepackage{amsfonts, tikz-cd}
\usepackage[T1]{fontenc}
\PassOptionsToPackage{linecolor=blue,backgroundcolor=blue!25,bordercolor=blue,textsize=scriptsize}{todonotes}
\usepackage{tikz}
\usetikzlibrary{backgrounds}
\usepackage[framemethod=tikz]{mdframed}
\AtBeginEnvironment{mdframed}{%
\tikzset{every picture/.style={}}%
}
\mdfsetup{roundcorner=.5ex}
{\end{mdframed}}

\DeclareSymbolFontAlphabet{\mathbb}{AMSb} 
\DeclareSymbolFontAlphabet{\mathbbl}{bbold}

\usepackage[sort&compress, comma, square, numbers]{natbib}
\usepackage[all,cmtip]{xy}
\usepackage{color}
\definecolor{MyDarkBlue}{rgb}{0.15,0.25,0.45}
\usepackage[linktocpage=true]{hyperref}
\hypersetup{colorlinks=true, citecolor=blue, linkcolor=blue, urlcolor=blue, pdfauthor={}, pdftitle={},breaklinks=true}

\let\SS=\S 

\newcommand{\dch}{{\check{\dd}}}
\newcommand{\Dch}{{\check{D}}}
\newcommand{\chF}{{\check{\mathcal F}}}

\newcommand{\chD}{{\check{D}}}

\renewcommand{\sp}{^{spur}}

\newcommand{\LC}{\text{\tiny LC}}
\newcommand{\CH}{\text{\tiny CH}}

\newcommand{\Bi}{{\text{\tiny B}}}

\renewcommand{\sb}{{\overline{\sigma}}}

\newcommand{\rb}{{\overline{ r}}}

\newcommand{\SO}{{\rm SO}}

\newcommand{\w}{{\,\wedge\,}}
\newcommand{\wt}{\widetilde}

\newcommand{\fD}{{\mathfrak{D}}}

\newcommand{\delslash}{\ensuremath \raisebox{0.025cm}{\slash}\hspace{-0.23cm} \del}
\newcommand{\Hslash}{\hspace{0.1cm}\ensuremath \raisebox{0.03cm}{\slash}\hspace{-0.30cm} H}
\newcommand{\Pslash}{\hspace{0.1cm}\ensuremath \raisebox{0.03cm}{\slash}\hspace{-0.28cm} P}
\newcommand{\Fslash}{\hspace{0.1cm}\ensuremath \raisebox{0.025cm}{\slash}\hspace{-0.28cm} F}

\newcommand{\half}{\frac{1}{2}}

\def\rep#1{{{\boldsymbol{#1}}}}

\def\CS{{\text{CS}}}

\newcommand{\bb}{{\overline\beta}}
\newcommand{\gb}{{\overline\gamma}}

\newcommand{\A}{\cA}

\newcommand{\X}{X}
\newcommand{\Y}{Y}
\newcommand{\B}{\ccB}
\newcommand{\Z}{\cZ}

\renewcommand{\aa}{\mathfrak{a}}

\renewcommand{\a}{\alpha}
\renewcommand{\b}{\beta}
\newcommand{\g}{\gamma}\newcommand{\G}{\Gamma}
\renewcommand{\d}{\delta}\newcommand{\D}{\Delta}
\newcommand{\e}{\epsilon}\newcommand{\ve}{\varepsilon}

\newcommand{\Th}{\Theta}

\renewcommand{\k}{\kappa}
\renewcommand{\l}{\lambda}\renewcommand{\L}{\Lambda}
\newcommand{\m}{\mu}
\newcommand{\n}{\nu}

\newcommand{\p}{\pi}
\renewcommand{\r}{\rho}
\newcommand{\s}{\sigma}\renewcommand{\S}{\Sigma}
\renewcommand{\t}{\tau}

\newcommand{\vph}{\varphi}

\renewcommand{\o}{\omega}\renewcommand{\O}{\Omega}

\DeclareFontFamily{OT1}{pzc}{}
\DeclareFontShape{OT1}{pzc}{m}{it}{<-> s * [1.200] pzcmi7t}{}
\DeclareMathAlphabet{\mathpzc}{OT1}{pzc}{m}{it}

\newcommand{\cA}{\mathcal{A}}
\newcommand{\ccB}{\mathpzc B}

\newcommand{\cD}{\mathcal{D}}

\newcommand{\cF}{\mathcal{F}}

\newcommand{\cL}{\mathcal{L}}

\newcommand{\cO}{\mathcal{O}}

\newcommand{\cR}{\mathcal{R}}

\newcommand{\ccY}{\mathpzc Y}
\newcommand{\cZ}{\mathcal{Z}}\newcommand{\ccZ}{\mathpzc Z}

\newcommand{\ccZb}{{\overline \ccZ}}

\DeclareFontFamily{U}{bbold}{}
\DeclareFontShape{U}{bbold}{m}{n}
 {  <-5.5> s*[1.05] bbold5
    <5.5-6.5> s*[1.05] bbold6
    <6.5-7.5> s*[1.05] bbold7
    <7.5-8.5> s*[1.05] bbold8
    <8.5-9.5> s*[1.05] bbold9
    <9.5-11.5> s*[1.05] bbold10
    <11.5-16> s*[1.05] bbold12
    <16-> s*[1.05] bbold17
 }{}

\renewcommand{\Im}{\mathbbl{m}}

\newcommand{\IR}{\mathbbl{R}}

\font\eightrm=cmr8 at 8pt
\font\csc=cmcsc10

\newcommand{\beq}{\begin{equation}}
\newcommand{\eeq}{\end{equation}}
\newcommand{\beqnn}{\begin{equation*}}
\newcommand{\eeqnn}{\end{equation*}}
\newcommand{\bea}{\begin{eqnarray}}
\newcommand{\eea}{\end{eqnarray}}
\newcommand{\bean}{\begin{eqnarray*}}
\newcommand{\eean}{\end{eqnarray*}}

\newcommand{\sref}[1]{\SS\ref{#1}}

\newcommand{\ii}{\text{i}}

\newcommand{\place}[3]{\vbox to0pt{\kern-\parskip\kern-7pt
                             \kern-#2truein\hbox{\kern#1truein #3}
                             \vss}\nointerlineskip}

\DeclareFontFamily{U}{wncy}{}
\DeclareFontShape{U}{wncy}{m}{n}{<->wncyr10}{}
\DeclareSymbolFont{mcy}{U}{wncy}{m}{n}
\DeclareMathSymbol{\sha}{\mathord}{mcy}{"58}

\newcommand{\del}{{\partial}}
\newcommand{\delb}{{\overline{\partial}}}

\newcommand{\lb}{{\overline\lambda}}
\newcommand{\nb}{{\overline\n}}
\newcommand{\mb}{{\overline\m}}

\newcommand{\Dbar}{{\overline D}}

\renewcommand{\aa}{\mathfrak{a}}
\newcommand{\mfb}{\mathfrak{b}}

\newcommand{\End}{{\text{End}\,}}

\newcommand{\dd}{{\text{d}}}
\newcommand{\ddn}{{\dd_\nabla}}

\newcommand{\ddH}{{\dd_{\nabla^\H}}}
\newcommand{\ddchH}{{\check\dd_{\nabla^\H}}}

\newcommand{\K}{K\"ahler\xspace}

\renewcommand{\H}{\text{H}}
\renewcommand{\B}{\text{B}}
\renewcommand{\Re}{\text{Re~}}
\renewcommand{\Im}{\text{Im~}}

\newcommand{\vol}{\mbox{\,vol}}
\newcommand{\tr}{\text{Tr}\hskip2pt}

\newcommand{\tb}{{\overline{\tau}}}

\newcommand{\ap}{{\a^{\backprime}\,}}
\renewcommand{\sb}{{\overline{\sigma}}}

\renewcommand{\rb}{{\overline{\rho}}}
\renewcommand{\=}{\;=\;}

\makeatletter
\g@addto@macro\bfseries{\boldmath}
\makeatother

\newcommand{\citeE}{\cite{McOrist:2021dnd}\xspace}
\newcommand{\citeES}{\cite{McOrist:2024zdz}\xspace}
\newcommand{\citeM}{\cite{Candelas:2016usb}\xspace}

\newcommand{\citeSG}{\cite{McOrist:2019mxh}\xspace}

\newcommand{\citeUG}{\cite{Candelas:2018lib, McOrist:2019mxh}\xspace}

\newcommand{\citeS}{\cite{Ashmore:2018ybe}\xspace}

\renewcommand{\baselinestretch}{1.1}
\numberwithin{equation}{section}
\proofmodefalse
\begin{document}
\pagestyle{empty}      
\ifproofmode\underline{\underline{\Large Working notes. Not for circulation.}}\else{}\fi

\begin{center}
\null\vskip-0.5in
{\Huge  The  physical 
moduli of heterotic $G_2$ \\string compactifications \\[0.3in]}
{\csc   Jock McOrist$^{\sharp 1}$,  Martin Sticka$^{\sharp 2}$ and Eirik Eik Svanes$^{\dagger 3}$\\[0.3in]}

{\it 
$^\sharp$ Department of Mathematics\hphantom{$^2$}\\
School of Science and Technology\\
University of New England\\
Armidale, 2351, Australia\\[3ex]
}

{\it 
$^\dagger$ Department of Mathematics and Physics\hphantom{$^1$}\\
Faculty of Science and Technology\\
University of Stavanger\\
N-4036, Stavanger, Norway\\[3ex]
}

\footnotetext[1]{{\tt jmcorist@une.edu.au}}
\footnotetext[2]{{\tt msticka@myune.edu.au}}
\footnotetext[3]{{\tt eirik.e.svanes@uis.no}}
\vspace{1cm}
{\bf Abstract\\[-15pt]}
\end{center}

In previous works, an operator was developed for heterotic compactifications on $\IR^{2,1}\times G_2$ and $AdS_3 \times G_2$, which preserves $N=1$ $d=3$ supersymmetry and whose kernel is related to the moduli of the compactification.  The operator is described in terms of non-physical spurious degrees of freedom, specifically, deformations of a connection on the tangent bundle. In this paper, we eliminate these spurious degrees of freedom by linking deformations of the spin connection to the moduli of the $G_2$ manifold $Y$. This results in an operator $\Dch$  that captures the physical moduli space of the $G_2$ heterotic string theory. When  $Y=X\times S^1$, with $X$ an $SU(3)$ manifold, we show $\Dch$ produces results that align with existing literature. This allows us to propose a $G_2$ moduli space metric. We check that this metric reduces to the $SU(3)$ moduli metric constructed in the literature. We then define an adjoint operator $\Dch^\dag$. We show the $G_2$ moduli correspond to the intersection of the kernels of $\Dch$ and $\Dch^\dag$.  These kernels reduce to the $SU(3)$ F-terms and D-terms respectively on $X\times S^1$. This gives two non-trivial consistency checks of our proposed moduli space metric.  Working perturbatively in $\ap$, we also demonstrate that the heterotic $G_2$ moduli problem can be characterised in terms of a double extension of ordinary bundles, just like in the $SU(3)$ case. 

\vskip150pt

\newgeometry{left=1.5in, right=0.5in, top=0.75in, bottom=0.8in}
%
\newpage
%
%
%
\setcounter{page}{1}
\pagestyle{plain}
\renewcommand{\baselinestretch}{1.3}
\null\vskip-10pt

\section{Introduction}
A heterotic compactification on a $G_2$-manifold $Y$ realises a three-dimensional spacetime
$$
\IR^{2,1} \times Y~, \qquad {\rm or} \qquad AdS_3 \times Y~,
$$
where the three-dimensional spacetime can be Minkowski or $AdS_3$. Such compactifications are potentially of interest to the $AdS_3/CFT_2$ correspondence, domain walls in four-dimensional spacetimes, or swampland conjectures. They were first studied in the context of compactifications in say \cite{Gunaydin:1995ku,Gauntlett:2001ur, Friedrich:2001nh, Friedrich:2001yp, Gauntlett:2002sc, Gauntlett:2003cy, Ivanov:2003nd, Ivanov:2009rh, Kunitomo:2009mx, Lukas:2010mf, Gray:2012md}.  Our focus is more formal; we aim to understand the moduli space of these compactifications. This moduli space is the manifold formed by the parameters describing unobstructed deformations of the equations of motion. The deformations we consider always preserve supersymmetry -- we currently lack the theoretical tools to consider deformations that break supersymmetry, though this remains of significant physical interest. Our primary tool is supergravity as an approximation to string theory. This should be viewed as a formal expansion in a small parameter $\ap\ll 1$ with a hierarchy of increasingly complex differential equations appearing order-by-order in $\ap$. The fields are similarly expanded in $\ap$, and the resulting system is solved order-by-order. We work correct to first order in $\ap$ in this expansion, where much is already known. The geometric data defining a solution to this order  includes the $G_2$ manifold $Y$, the $G_2$ structure $\varphi$, a compatible metric $g$ on $Y$, a bundle $V \to Y$ with a connection $A$ whose structure group embeds in $E_8 \times E_8$, and a three-form $H$. The supersymmetry conditions imply $F$ satisfies an instanton condition \cite{Gauntlett:2002sc, Gauntlett:2003cy,Ivanov:2003nd,ReyesCarrion:1998si}:
$$
F \w \psi \= 0~.
$$
How this collection of equations and resulting solutions are modified at higher orders in the expansion in $\ap$ is interesting. While we do not know what they are explicitly, we at least need the corrections to be smaller than the first order analysis discussed here. Returning to the first order calculation, to be of interest to heterotic compactifications the $G_2$ structure $\varphi$ must be integrable \cite{fernandez1998dolbeault, Gauntlett:2001ur, Friedrich:2001yp} and in this situation there is  a unique metric connection on the tangent bundle $TY$ with totally antisymmetric torsion \cite{Friedrich:2001nh}, we call it $\Th^B$. There are two other  relevant connections: the Hull connection $\Th^\H$ which has completely antisymmetric torsion of opposite sign to $\Th^B$, and the Levi-Civita connection $\Th^\LC$. We denote these curvatures $R^\H$, $R^B$, and  $R$ for Levi-Civita. These connections are all determined in terms of the underlying metric and three-form $H$.  
This data satisfies a Bianchi identity that derives from the Green-Schwarz anomaly cancellation condition on the worldsheet:
\beq
\dd H \=\!- \frac{\ap}{4} \Big( \tr (F\w F) - \tr ( R^\H\w R^\H)\, \Big)~.
\label{eq:Anomaly0}
\eeq
The connection used in $\tr R^\H \w R^\H$ is determined by matching with string theory scattering amplitudes. It is also the connection preferred by a compatibility between the supergravity equations of motion and supersymmetry. For this reason, it is the moduli space of the string theory defined by this connection that we wish to study.

We are interested in the space of deformations of these equations, known as the moduli space. This has been studied recently in \cite{delaOssa:2016ivz, Clarke:2016qtg, delaOssa:2017pqy, delaOssa:2017gjq, Fiset:2017auc, Clarke:2020erl, Silva:2024fvl}. The moduli of string compactifications with $H=0$ were studied in \cite{delaOssa:2016ivz} in which $Y$ has $G_2$ holonomy. This is the analogue of the standard embedding for $SU(3)$ compactifications. It was found in that paper that the moduli space obeyed an Atiyah type equation, in which deformations of the $G_2$ structure were obstructed by the presence of the bundle. As in the case of Atiyah, the problem is a description of simultaneous deformations of complex manifolds with a holomorphic bundle, and some of these deformations are potentially obstructed in that the bundle could generate some $F^{0,2}$. This condition is captured by a matrix operator on sections of a certain bundle, and its kernel is the space of deformations. In the case of $G_2$ holonomy, a similar matrix structure and outcome was developed in \cite{delaOssa:2016ivz}. See also \cite{Anderson:2010mh, Anderson:2011ty} for applications of this mechanism to the heterotic $SU(3)$ system. 

There is an important result \cite{Clarke:2016qtg} that should be highlighted. It was proven mathematically that by allowing the connection on the tangent bundle to vary independently of the other moduli, that the space of deformations of $N=1$ $d=3$ is in fact finite dimensional up to and including the first order $\ap$--correction. As the space of physical deformations is a sub-locus of this finite dimensional space, assuming everything is smooth, the space of physical deformations is also finite dimensional. This is an important result especially for the following physical reasons. Recall that for $N=1$ $d=4$ heterotic vacua,  \cite{Nemeschansky:1986yx,Jardine:2018sft} and \cite{Witten:1985bz,Witten:1986kg} who show that from the worldsheet and spacetime point of view respectively that the moduli of $N=1$ $d=4$ vacua exist and are finite dimensional even once all the modifications in the equations of motion and fields are included,\footnote{See also \cite{Garcia-Fernandez:2015hja} for a proof that the moduli space of the formally defined Hull-Strominger system is finite dimensional, where again the connection on the tangent bundle is allowed to vary independently of the other moduli.} no such proof exists for $N=1$ $d=3$ theories. In fact, \cite{Becker:2014rea} point out that the expectation is in fact that there are generically no vacua: the $\ap$-corrections are expected to lift the vacua so that there are no solutions. While the proof of \cite{Clarke:2016qtg} that if the vacua exist the moduli is finite dimensional (and so gives us hope they exist in the first place), we really need to understand the role of the $\ap$-corrections, beyond the first order case, for $N=1$ $d=3$. For our purposes however we do not deal with this question but it would be interesting to pursue in the future.

This work was extended to the case of flux in \cite{delaOssa:2017pqy}. This requires dealing with a non-trivial Bianchi identity and recently explicit solutions with $AdS_3$ spacetimes have been constructed in \cite{delaOssa:2021qlt, Lotay:2021eog, Galdeano:2024fsc}. Our interest is in Minkowski spacetime however. Indeed, the natural direction to take is inspired by the $SU(3)$ analysis wherein one defines a complex of vector bundles with an operator. In contrast to $SU(3)$ structure manifolds, $G_2$ is odd-dimensional and so there is a single $D$-operator whose nilpotentency relies explicitly on the representation theory of $G_2$, as well as the Bianchi identites for the curvatures $R^\H, F$ and $H$. That being so, we will often have need to explicitly project onto irreducible representations, and so in the next section we review some of the relevant formalism. Indeed, integrable $G_2$ structures have features in common with complex geometry, hence the name integrable. A canonical differential complex $\Lambda^*(Y)$ can be defined as a sub-complex of the de Rham complex and the associated cohomologies $H^*(Y)$  have similarities with the Dolbeault complex of complex geometry. Families of $SU(3)$ structure manifolds can be examined through an embedding in integrable $G_2$ geometry, see e.g. \cite{delaOssa:2014lma}. This embedding allows variations of complex and Hermitian structures of six-dimensional manifolds to be considered equivalently.

In \cite{delaOssa:2017pqy} the moduli space of the supersymmetric $G_2$ solutions is studied by modifying the string theory Bianchi identity \eqref{eq:Anomaly0} so that the connection used to compute $\tr R^2$ is an instanton with respect to the $G_2$ structure.  While the Hull connection is an instanton to zeroth order in $\ap$, the instanton connection carries with it additional non-physical deformations. These introduce non-physical or spurious directions in the moduli space and, following \citeE, our goal here is to study the deformation theory of the $G_2$ compactifications without the spurious modes. Our approach is to directly utilise that the Hull connection is actually defined in terms of the metric and $H$.  We find an operator $\check D$ that squares to zero to the appropriate order in $\ap$. An important check is that when we dimensionally reduce by for example an ans\"atz $Y=X\times S^1$ where $X$ is $SU(3)$ structure, we find that $\check D$ reduces to a pair of operators $\Dbar$ and $D$. These operators were constructed in \citeE and \citeES respectively, and were shown to capture the F-terms and D-terms of $SU(3)$ compactifications preserving $N=1$, $d=4$ supersymmetry. This is a non-trivial consistency check of both the work in this article as well as the works in \citeE and \citeES. We give a conjecture for the $G_2$ moduli space and evidence for this is that upon reduction to $SU(3)$  reduces to the result \citeSG. The $G_2$ moduli metric allows us to construct an adjoint operator $\Dch^\dag$. The kernel of this operator corresponds to gauge fixing in $G_2$ but upon reduction, with an appropriate choice of holomorphic coordinates, matches the D-terms of the $SU(3)$ theory. Together these form a non-trivial consistency check of the moduli space metric including its $\ap$-corrections. Finally, we observe that by formally separating the components of the deformations in an $\ap$ structure, we can redefine the operator $\check D$ so that it is upper triangular. This could lead to a new understanding of the $G_2$ moduli, as the upper triangular structure facilitates building a complex in terms of short exact sequences, which in turn allows one to compute the cohomology of $\check D$ using homological algebra techniques, just as in the $SU(3)$ case \cite{Anderson:2010mh, Anderson:2011ty, Melnikov:2011ez, Anderson:2014xha, delaOssa:2014msa, McOrist:2021dnd}.

\section{Supersymmetric heterotic supergravity solutions}

Connections with torsion are central to our discussion. The literature is a veritable banquet of varying notation and conventions, so we verbosely lay out our approach in the interest of disambiguation. In appendix \sref{app:literature} we provide a dictionary for translation with some other key papers. We hope this will be of a modicum of help to the community. 

Suppose we have a Riemannian manifold $Y$ with real coordinates $x^m$ and a metric $g_{mn}$. On the tangent bundle we define a covariant derivative. In the coordinate basis the connection symbols are defined as
\beq
\nabla_m \del_n \= \G_{m}{}^p{}_n \del_p~, \qquad \nabla_m \dd x^n \=\! - \G_{m}{}^n{}_p \dd x^p~. 
\eeq
To deal with spinors we endow the manifold with a vielbein $e^a = e^a_m \dd x^m$, its inverse $E_a = E_a^m \del_m$ and hermitian gamma matrices satisfying a Clifford algebra $\{ \g^a, \g^b\} = 2 \eta^{ab}$, where $\eta_{ab} = {\rm diag} \{1,1,\cdots,1\}$ is the flat metric acting on the tangent space indices; it is related to the curved metric via the vielbein $g_{mn} = e_m^a e_n^b \eta_{ab}$.  In the vielbein basis we define a spin connection $\Th_{ab}  =\dd x^m \Th_{mab} $ as
\beq
\nabla_m e^a \=\!- \Th_m{}^a{}_b e^b~, \qquad \nabla_m E_a \= \Th_m{}^b{}_a E_b~,
\eeq
which appears in the gravitino variation, and defines a covariant derivative
 \beq
\nabla_m \e \= \del_m \e + \frac{1}{2} \Th_{m\,ab} \S^{ab} \e~,
 \eeq
 where $\S^{ab} = \frac{1}{4} [\g^{a},\g^b]$ are anti-hermitian generators of $\SO(d)$ with $\g^a$ hermitian gamma matrices. The connection being metric compatible means $\Th_{mab} = - \Th_{mba}$. The torsion $T$ and curvature $R$ can be computed from the Cartan structure equations (e.g. \cite{Eguchi:1980jx}):
 \beq\label{eq:cartan1a}
\begin{split}
  \dd e^a + \Th^a{}_b e^b &\= T^a \= \half T^a{}_{bc} e^b \w e^c~,\\
  \dd \Th^a{}_b + \Th^a{}_c \w\Th^c{}_b &\= R^a{}_b \= \half  R^a{}_{bcd}  e^c \w e^d~.
\end{split}
 \eeq
The torsion of $\Th$ as defined above is not the same as the contorsion -- the difference between the connection of interest and the Levi-Civita connection. It depends on conventions and definitions. In our case, we are interested in torsion that is completely antisymmetric, and the two differ by a sign. The  two relevant cases are
\beq\label{eq:OmTorCurv}
\begin{split}
 \Th^\H_{mab} &\= \Th_{mab}^{\LC} + \half H_{mab}~, \qquad T^{\H\,a}{}_{bc} \=\! - H^a{}_{bc}~,\\
 \Th^B_{mab} &\= \Th_{mab}^{\LC} - \half H_{mab}~, \qquad T^{\B \,a}{}_{bc} \=  H^a{}_{bc}~,\\
\end{split}
\eeq
We have used \eqref{eq:cartan1a} to compute the torsion.  The spin connection is related to the connection symbols that define covariant derivatives on tensors\footnote{The indices in $\G_m{}^n{}_{p}$ appear in every possible order in the literature. We have chosen a convention but the readers mileage might vary.  The order of indices is not physically important, but once a convention is set for $\Th$ and $\G$ it is important to stay true to that convention. Given a convention, to derive the relation to $\Th_m$ its straightforward to use the definitions of $\G$ and $\Th$:
 $$
 \nabla_m \del_n= \G_m{}^p{}_{n} \del_p~, \qquad \nabla_m E_a = \Th_m{}^b{}_a E_b~,
 $$ and then apply the transformation to the vielbein e.g. $\del_p = e_p^b E_b$. }
\beq\label{eq:GammaOm}
 \G_m{}^n{}_{p} \= E^n_a \, (\Th_m{}^a{}_b )\, e_p^b + E_c^n\, \del_m e^c_p~.
\eeq
The advantage of this convention is that the translation from $\Th$ to $\G$ is obvious:
$$
\Th^\H_{mab} \= \Th_{mab}^{\LC}  +\half  H_{mab}~, \qquad \G^\H_{m}{\,}^p{}_q  \= \G^\LC_{m}{\,}^p{}_q  + \half H_m{}^p{}_q~. \qquad
$$
As an example, $\nabla^\H$ acting on a vector is
\beq\label{eq:covderivV}
\nabla^\H_m V^n \=  \nabla^\LC_m V^n + \half H_m{}^n{}_{p}  V^p~.
\eeq
An equivalent definition of torsion in terms of a coordinate basis is
\beq\label{eq:torsion2}
\begin{split}
 T \= \nabla_m \del_n - \nabla_n \del_m \= \left(\G_m{}^p{}_{n} - \G_n{}^p{}_{m}\right) \del_p~.
\end{split}
\eeq
One can check that applying this definition of torsion to \eqref{eq:covderivV} gives the same result as the Cartan structure equation \eqref{eq:OmTorCurv}. This is not surprising --- Cartan's formulation is equivalent to the conventional coordinate description and the map \eqref{eq:covderivV}, which is formally the same as a gauge transformation, does not modify physical quantities like torsion. 

A second straightforward calculation relevant to supersymmetric compactification is to posit the existence of a globally well-defined spinor $\ve$ such that
\beq
\nabla^B \ve \= \nabla^\LC \ve -\frac{1}{8} H_{mab} \g^{ab} \ve \= 0~,
\eeq
and from this construct a tensor via a spinor bilinear. An example is a vector such as $V^m = \ve^\dag \gamma^m \ve$.\footnote{There are details such as the overall normalisation of the bilinear that we are suppressing as they depend on the particular dimension of the manifold $Y$ and the reality properties of $\ve$; those overall constants are not important right now.} The calculation is spelled out in the appendix, one finds the same result as  \eqref{eq:GammaOm}:
$$
\nabla^B_m V^n \=  \nabla^\LC_m V^n  - \half H_m{}^n{}_{p}  V^p~.
$$

\section{\texorpdfstring{$G_2$}{G2} compactifications of heterotic string theory}

\subsection{Ten-dimensional supergravity with first order \texorpdfstring{$\ap$}{alphaprime}: conventions}
In the field basis of \cite{Bergshoeff:1989de,Bergshoeff:1988nn} the ten--dimensional $\ap^2$--supergravity theory can be written down explicitly. In the Appendix we give a translation of notation from those papers to the convention used here. The action is
\begin{equation}
S \= \frac{1}{2\kappa_{10}^2} \int\! \dd^{10\,}\! X \sqrt{g_{10}}\, e^{-2\Phi} \Big\{ \cR -
\half |H|^2  + 4(\del \Phi)^2 - \frac{\alpha'}{4}\big( \tr |F|^2 {-} \tr |R^\H|^2 \big) \Big\} + \cO(\alpha'^3)~.
\label{eq:10daction}
\end{equation}
On the $G_2$ manifold $Y$, we denote $m,n,\ldots$ real indices along $Y$.\footnote{The 10D Newton constant is denoted by $\kappa_{10}$,
\hbox{$g_{10}=-\det(g_{MN})$}, $\Phi$ is the 10D dilaton.} $\cR$ is the Ricci scalar evaluated using the Levi-Civita connection and $F$ is the Yang--Mills field strength with the trace taken in the adjoint of the gauge group. We take the generators of the gauge group $T^a$ and the trace $\tr$ to be defined so that  $\tr (T^a T^b)$ is positive definite. We will often need wedge products involving the                        basis elements of $T^*Y$ written in a coordinate basis. To save writing excessive wedges we will write $\dd x^{m_1\cdots m_n} \cong \dd x^{m_1} \w \cdots \wedge \dd x^{m_n}$. If any possible confusion arises we will use the $\wedge$ symbol explicitly. 

There is a point-wise inner product  on $p$-forms given by
$$
\langle S,\, T\rangle~=~ 
\frac{1}{p!} \, g^{M_1 N_1} \ldots g^{M_p N_p}\, S_{M_1\ldots M_p} \,T_{N_1 \ldots N_p}~, \qquad |T|^2 ~=~\langle T,\, T\rangle~.
$$
 The three--form $H$ satisfies a Bianchi--identity
\beq\notag
\dd H \=\!- \frac{\ap}{4} \left( \tr F\w F - \tr R^\H\w R^\H \, \right)~.
\eeq

The 10-dimensional background being supersymmetric amounts to the vanishing gravitino, dilatino and gaugino variations:
\beq
\begin{split}
 \d \Psi_M &\= \nabla_M^B \ve  + \cO(\ap^2) \= 0 ~,\\
 \d \lambda &\=\! - \frac{1}{2\sqrt{2}} \left( \delslash - \half \Hslash  \right) \ve  + \cO(\ap^2) \= 0~,\\
 \d \chi &\=\! - \frac{1}{2} \Fslash \ve + \cO(\ap^2) \= 0~,
\end{split}
\label{eq:10dsusy}
\eeq
where we use the usual slash notation. We have written the formal expansion parameter $\ap^2$ explicitly; there are equations not written to do with second order equations, beyond our remit. The important point is that the spinor $\ve$ is covariantly constant with  torsion opposite to what appears in the action, Bianchi identity and equations of motion.

\subsection{\texorpdfstring{$G_2$}{G2} manifolds}
First,  we briefly recapitulate a $G_2$ compactification of string theory using the   formalism described in \cite{delaOssa:2017pqy}. This will also serve as a way of establishing our notation. We now suppose that the spacetime is $\IR^{2,1}\times Y$ with $Y$  a 7-dimensional Riemannian manifold. 
 The existence of a covariantly constant spinor \eqref{eq:10dsusy} implies the existence of a $G_2$-structure 3-form $\phi$ constructed as a spinor billinear. The form $\phi$ determines a Riemannian $G_2$ metric $g$, and in virtue of \eqref{eq:10dsusy} both are covariantly constant
\beq\notag
\nabla^B g \= 0~, \qquad \nabla^B \phi \= 0~.
\eeq
There is an associated four-form $\psi =\star\,\phi$. The manifold $Y$ has $G_2$ holonomy if and only if $\dd \phi =0$ and $\dd \psi = 0$.  We require that $Y$ instead has an integrable $G_2$ structure which amounts to the $\phi,\psi$ no longer being closed but that there is the vanishing of a particular torsion module. Given this, we will be able to define an analogue of a Dolbeault operator  $\dch$ for complex manifolds. 

We divide up our forms according to type. Denote $\L^k(Y)$  the space of $k$-forms on $Y$. To arrive at a differential operator that replicates Dolbeault cohomology it is important to project onto representations of $G_2$. We denote  $\L^k_{\rep{p}}(Y)$  the subspace of $k$-forms which transform in the $\rep{p}$ representation of $G_2$, and  $\pi_{\rep{p}}:\L^k(Y) \to \L^k_{\rep{p}}(Y)$ the projection of a $k$-form on to the representation $\rep{p}$  of $G_2$. On $Y$ therefore, we have a decomposition of forms given as
\beq\label{eq:forms}
\begin{split}
& \L^0 \= \L_{\rep{1}}^0~,\\
& \L^1 \= \L_{\rep{7}}^1 = T^*Y \cong TY~,\\
& \L^2 \= \L_{\rep{7}}^2  \oplus \L^2_{\rep{14}}~,\\
& \L^3 \= \L_{\rep{1}}^3  \oplus \L^3_{\rep{7}} \oplus \L^3_{\rep{27}}~,
\end{split}
\eeq
the remainder taken care of by the Hodge dual with respect to $g$. The explicit projections used to decompose the forms in \eqref{eq:forms} into irreducible representations are summarised in Appendix \sref{s:projections}.

To construct the complexes required for the discussion of the moduli, we need to define an analogue of the Dolbeault operator for $G_2$ manifolds. We define an operator $\dch$ given by
\beq
\begin{split}
& \dch_0: \L^0(Y) \to \L^1(Y)~, \qquad \dch_0 f \= \dd f~, \qquad~~ f\in \L^0(Y)~,\\
& \dch_1: \L^1(Y) \to \L_{\rep{7}}^2(Y)~, \qquad \dch_1 \a \= \pi_7(\dd \a)~, \quad \a\in \L^1(Y)~,\\
& \dch_2: \L_{\rep{7}}^2(Y) \to \L_{\rep{1}}^3(Y)~, \qquad \dch_2 \b \= \pi_1(\dd \b)~, \quad \b\in \L^2(Y)~.
\end{split}
\eeq
If we consider a series of maps generated by $\dch$:
\beq
0 \rightarrow \L^0(Y) \xrightarrow{\dch}  \L^1(Y) \xrightarrow{\dch}  \L^2_{\rep{7}} (Y) \xrightarrow{\dch}  \L^3_{\rep{1}}(Y) \rightarrow 0~,\label{eq:dchcomplex}
\eeq
then this is a complex,  $\dch^2 = 0$, if and only if the $G_2$ structure is integrable. In this paper, we take the $G_2$ structure to be integrable so that \eqref{eq:dchcomplex} is indeed a complex. We denote the complex $\check\L^* (Y)$ and it is in fact elliptic with the corresponding cohomology ring $\check H^*(Y)$ being the canonical $G_2$-cohomology of $Y$.

The complex has an extension to bundle valued forms in a natural way.  We define
\beq
\begin{split}
& \dch_{0\,A}: \L^0(Y,E) \to \L^1(Y,E)~, \qquad \dch_{0\,A} f \= \dd_A f~, \qquad~~ f\in \L^0(Y,E)~,\\
& \dch_{1\,A}: \L^1(Y,E) \to \L_{\rep{7}}^2(Y,E)~, \qquad \dch_{1\,A} \a \= \pi_7(\dd_A \a)~, \quad \a\in \L^1(Y,E)~,\\
& \dch_{2\,A}: \L_{\rep{7}}^2(Y,E) \to \L_{\rep{1}}^3(Y,E)~, \qquad \dch_{2\,A} \b \= \pi_1(\dd_A \b)~, \quad \b\in \L^2(Y,E)~.
\end{split}
\eeq
We have that $\dch_A^2 = 0$ if and only if the connection on $A$ is an instanton and the manifold has integrable $G_2$ structure. Instanton in the context of $G_2$ manifolds means 
$$
F \w \psi \= 0~. 
$$

The heterotic Bianchi identity is
\beq
\dd H \= - \frac{\ap}{4} \left( \tr F \w F - \tr  R^\H \w  R^\H \right)~,\label{eq:Bianchi}
\eeq
where from the gravitino variation we find the $G_2$ structure satisfies
$$
\dd \vph \=\!- \frac{1}{4} H_{mn}{}^p \vph_{prs} \dd x^{mnrs}~, \qquad \dd \psi \=\!- \frac{1}{12} H_{mn}{}^p\psi_{prst} \dd x^{mnrst}~,
$$
where we have assumed the $G_2$ structure integrable.

As a first step we follow \cite{delaOssa:2016ivz,delaOssa:2017pqy} by introducing the spurious modes, corresponding to deformations of the connection on $TY$, and define an operator that will hand us a complex.  The bundle $Q\sp$ is defined as 
$$
Q\sp\= T^*Y \oplus {\rm End}(TY) \oplus {\rm End}(V)~.
$$
A section  is denoted by
$$
\ccY\sp \= 
\begin{pmatrix}
 M \\
 K\\
 \aa
\end{pmatrix}~.
$$
We will need to study $p$-forms valued in $Q\sp$ and by an abuse of notation use the same symbol for such a section: $\ccY\sp\in \L^p (Y, Q)$. In calculations we will have need for indices, and the order of the indices is important as they are not democratic. For example, $M \cong M_{P m} \dd x^P \otimes \dd x^m$, where $P$ is a multi-index of degree $d$ with $\dd x^P = \dd x^{p_1\cdots p_d}/d!$. Very often $d=1$ and so $M$ is a 1-form valued in the cotangent bundle $T^*Y$. We sometimes explicitly write the index corresponding to the cotangent bundle viz. $M_m\cong M_{Pm} \dd x^P$.

We define an operator on $Q\sp$-valued forms:
\beq\label{eq:D}
D\sp:\L^p(Y,Q) \to \L^{p+1}(Y,Q\sp)~, \qquad \qquad D\sp \= 
 \begin{pmatrix}
\dd_{\nabla^\H} &  \cR_1 & -\cF_1\\
  \cR_2 & \dd_{ \nabla^\H} & 0\\
 \cF_2 & 0 & \dd_A
\end{pmatrix}~,
\eeq
where
\beq\label{eq:Dspur2}
\begin{split}
  \cF_1(\aa)_m &\=   \frac{\ap}{4} \tr (  F_{mq} \dd x^q \w \aa)~,\qquad \cF_2(M) \= g^{mn} F_{nq} M_{Pm} \dd x^q \w   \dd x^{P} ~, \\
 \cR_1(K)_m &\=   \frac{\ap}{4} \tr ( R^\H_{mq} \dd x^q \w K) ~, \qquad \cR_2(M)\=  g^{mn}  R^\H_{nq}  M_{Pm} \dd x^q \w \dd x^{P}~.
\end{split}
\eeq

This operator $D\sp$ does not, in fact, square to zero as written. Instead, we construct a related operator $\check D\sp$ defined by projecting   $D\sp$ component-wise onto certain irreducible representations of $G_2$ in analogy with $\dch, \dch_A$. For example, for  $\cF_2$ we define
$$
\chF_2:\L_{\rep{r}}^{(p)}(Y,TY) \quad \longrightarrow \quad \L_{\rep{r}'}^{(p+1)} (Y, {\rm End}(V))~,
$$
with
\beq
\begin{split}
 \chF_2(M) &\= \cF_2(M)  = M^m F_m~, ~~\qquad\qquad\qquad M \in \L^0(TY)~,\\
 \chF_2(M) &\= \pi_{\rep{7}} (\cF_2(M))  = -\pi_{\rep{7}} (M^m \w F_m)~, \quad\,\, M \in \L^1(TY)~,\\
 \chF_2(M) &\= \pi_{\rep{1}}(\cF_2(M))  =\pi_{\rep{1}} ( M^m \w F_m)~, ~~\quad\,~ M \in \L_{\rep{7}}^2(TY)~.\\
\end{split}
\eeq
There are similar definitions for $\cR_i, \cF_i$. The end result is 
\beq\label{eq:Dch}
\check D\sp:\L_{\rep{r}}^p(Y,Q) \to \L^{p+1}_{\rep{r}'}(Y,Q)~, \qquad \qquad \check D\sp \= 
 \begin{pmatrix}
\dch_{\nabla^\H} &  \check \cR_1 & - \check \cF_1\\
  \check \cR_2 & \dch_{ \nabla^\H} & 0\\
 \check \cF_2 & 0 & \dch_A
\end{pmatrix}~.
\eeq

The operator $\Dch\sp$ squares to a diagonal matrix after using the Bianchi identities for $F$, $R^\H$ as well as $H$ in \eqref{eq:Bianchi} together with the relation between Hull and Bismut curvatures
\beq
 R^\H_{mnab} - R^\Bi_{abmn}  \= \half (\dd H)_{mnab}~.
\eeq
It in fact squares to zero after projecting onto the appropriate representations, noting the curvatures satisfy
\beq
\psi \w R^B_{abmn}\dd x^{mn} \= \cO(\ap^2)~, \quad \psi \w R^\H_{mnab}\dd x^{mn} \= \cO(\ap)~.
\eeq
If the $G_2$ manifold is reduced via $Y=X\times S^1$, where $X$ is an SU(3) manifold, the first condition above amounts to the metric on $X$ being Bismut Ricci-flat \cite{McOrist:2024zdz}, and Bismut Ricci-flat is the same as conformally balanced. We return to this reduction later.

This means we can define a complex 
\beq
0 \rightarrow \L^0(Y,Q\sp) \xrightarrow{\Dch\sp}  \L^1(Y,Q\sp) \xrightarrow{\Dch\sp}  \L^2_{\rep{7}} (Y,Q\sp) \xrightarrow{\Dch\sp}  \L^3_{\rep{1}}(Y,Q\sp) \rightarrow 0~,\label{eq:DchcomplexSpur}
\eeq
which is in fact elliptic. 

The heterotic moduli in \cite{delaOssa:2017pqy} are described by the kernel of the $\Dch\sp$-operator, and they include the spurious degrees of freedom, labelled here by $K$:
\beq\label{eq:DYspur}
\Dch\sp \ccY\sp \= 
\begin{pmatrix}
 \ddchH M + \check \cR_1(K) - \check \cF_1(\aa)\\
  \ddchH  K +  \check\cR_2(M)\\
 \check\dd_A \aa + \check \cF_2(M)
\end{pmatrix}\=0~. 
\eeq

The moduli problem can be phrased in terms of a connection on the bundle $Q$:
$$
\cD = \dd_\A \= \dd + \A~,
$$
and $\cD^2 = 0$ is equivalent to 
$$
F(\A) \w \psi \= 0~, \qquad F(\A) = \dd \A + \A \w \A~,
$$
so that $\cD$ defines an instanton connection on $Q\sp$ if we have a solution of the heterotic supersymmetry equations. This observation was in fact the  inspiration for the work in \citeES. 

In the next section we will construct a bundle $Q$ and an operator $\Dch$ without the spurious degrees of freedom.

\section{Constructing the \texorpdfstring{$G_2$}{G2}--complex for physical moduli}
\label{sec:DiffComp}
Let us assume $H = \cO(\ap)$ so that  $Y$ is $G_2$ holonomy to zeroth order in $\ap$ and the corresponding curvature is Ricci-flat. This is  a physical  requirement -- if we take $H=\cO(\ap)$, it is guaranteed that the $G_2$ geometry will define a weakly coupled worldsheet sigma model that underpins a string theory vacuum. If we do not make this assumption, the status of the compactification is murkier as far as quantum corrections go --- that is, do the solutions to the higher order $\ap$ equations  start to dominate the zeroth and first order solution? Indeed,  there are  $SU(3)$   heterotic compactifications defined by string theory dualities e.g. \cite{McOrist:2010jw}, but these do not have sigma model descriptions, and so it is unclear at best to what extent the supergravity approximation is valid for such backgrounds. That being so, we study the case $H=\cO(\ap)$.  

We want to eliminate the spurious term, which amounts to the component labelled by $K$ in the previous section, following the calculation in  \citeE. To that end consider  \eqref{eq:DYspur} acting on a 1-form valued in $Q$. The first  equation  is
\beq\label{eq:Deq1}
\psi\w \left( \ddH M_m  + \frac{\ap}{4} \tr \left(\aa \w F_m \right) - \frac{\ap}{4} \tr \left( K \w  R^\H_m \right)\right) \= 0~,
\eeq
where whenever an index is written explicitly we take the derivative to act on it as the Hull connection, and if it is suppressed as a form index it acts like the Levi-Civita connection.  For example, from \eqref{eq:DYspur}
\beq\label{eq:Deq2}
 \ddH M_m \=  \left(\del_p M_{qm}  - \G^\H_p{}^s{}_m M_{qs}\right) \dd x^{pq}~, \quad \frac{\ap}{4} \tr K \w R_m^\H \= \frac{\ap}{4} R_{pm}{}^s{}_t  K_q{}^t{}_s \, \dd x^{pq}~.
\eeq
As the Levi-Civita connection is torsionless, it drops out when the wedge product is used for the derivative, as per the case above. The projection onto the $\rep{7}$ is achieved by wedging with $\psi$. The calculation is to first order in $\ap$ and so we drop the Hull superscript on $R$ and take it to be Levi-Civita.  Our goal is to rewrite the last term of \eqref{eq:Deq1} in terms of the underpinning $G_2$ moduli.

We  use that $K$ is a deformation of the Hull connection. We compute first
\beq\label{eq:defOfHull}
\d \G^\H_q{}^s{}_t \= \d \G^\LC_q{}^s{}_t + \half \d H_q{}^s{}_t ~.
\eeq
A deformation of the Levi-Civita symbol $\G^\LC$ and three-form $H$ is
\beq\label{eq:deltaH}
\begin{split}
\d \G^\LC_q{}^s{}_t &\= g^{sp} \left( \nabla_{[t} \d g_{p]q} + \half \nabla_q \d g_{tp} \right)~,\\
\d H_q{}^s{}_t &\= g^{sm} \d g_{mn} g^{np} H_{qtp} + g^{sp}\left( \nabla_q \ccB_{pt} + \nabla_p \ccB_{tq} + \nabla_t \ccB_{qp} \right)~,
\end{split}
\eeq
where 
$$
\d H \=  \dd \ccB - \frac{\ap}{2} \tr \big( \d A F \big)+ \frac{\ap}{2} \tr \big( K R \big)~,
$$
 with $\ccB$ a gauge invariant combination of the $B$-field and gauge connection 
 $$
 \ccB  \=  \d B +\frac{\ap}{4}\tr{(A\,\d A)} - \frac{\ap}{4}\tr{(\Th\,K )} + \dd \textrm{-closed}~.
 $$
 Combining \eqref{eq:deltaH} to \eqref{eq:defOfHull}, and then transforming $\d \G$ to a deformation of the spin connection $K$ (see e.g. \cite{Eguchi:1980jx}), we find 
\beq
\frac{\ap}{4} \tr K \w  R_m  \= \frac{\ap}{4} R_{pm}{}^{ts} K_q{}_{st} \dd x^{pq} = \frac{\ap}{4} R_{pm}{}^{ts} \nabla_t (\d g + \ccB)_{qs} \dd x^{pq} ~.
\eeq
Setting 
\beq\label{eq:M}
M_{ab} = \frac{1}{2} (\d g_{ab} + \ccB_{ab})  + \cO(\ap)~,
\eeq
 we have
\beq
\frac{\ap}{4} \tr K \w  R_m \= \frac{\ap}{2}  \left(  R_{pm}{}^{ts} \nabla_t M_{qs} \right)\dd x^{pq}~.
\eeq
 We substitute this into \eqref{eq:Deq1},
\beq\label{eq:Moduli1}
\psi\w \left( \ddH M_{m}  + \frac{\ap}{4} \tr \left( \aa \w F_m \right) - \frac{\ap}{2} R_{m}{}^{ts} \nabla_t M_s \right)\= 0~,
\eeq
where $F_m = F_{mq} \dd x^q$, $R_m{}^t{}_s = R_{qm}{}^t{}_s \dd x^q$, and $M^s = M_q{}^s \dd x^q$ and  $\psi \w$ projects onto the $\rep{7}$. Recall that $\nabla^\H$ in \eqref{eq:Moduli1} acts as the Hull connection on the indices corresponding to  $T^*Y$, and as Levi-Civita on the form indices; while $\nabla$ is the Levi-Civita connection, but due to the overall $\ap$ could be taken to be Hull, Chern or Bismut.

The end result is that we define a bundle $Q$, which is topologically 
$$
 Q\= T^*Y \oplus {\rm End}(V)~,
$$
with sections corresponding to moduli
\beq\label{eq:Y}
\ccY \= 
\begin{pmatrix}
 M \\
 \aa
\end{pmatrix}~.
\eeq
This is acted upon by an operator
\beq\label{eq:Dtilde}
 \cD \= 
 \begin{pmatrix}
 \ddH + R \nabla   & - \cF_1\\
 \cF_2  & \dd_A
\end{pmatrix}~,
\eeq
where 
\beq\label{eq:defnRnabcF}
\begin{split}
 (R\nabla)( M_m)& \= - \frac{\ap}{2} \, R_{qm}{}^{ts} \nabla_t  M_{P\,s}\, \dd x^q\w \dd x^P~,  \\
\cF_1(\aa)_m &\=   \frac{\ap}{4} \tr (  F_{mq} \dd x^q \w \aa)~,\qquad \cF_2(M) \=  g^{mn}  F_{nq} M_{Pm} \,\dd x^q\w \dd x^P ~.
\end{split}
\eeq

\subsection{ Demonstrating \texorpdfstring{$\chD^2=0$}{D2=0} }
\label{sec:chDNil}
We now demonstrate that $\chD$ satisfies $\chD^2 = 0$. It is an important consistency check of our previous calculation. 

Schematically, the square of \eqref{eq:Dtilde} is
\beq\label{eq:DtildeSquared}
\cD^2 \= 
 \begin{pmatrix}
(\ddH + R \nabla)^2 - \cF_1 \cF_2   & - (\ddH + R \nabla) \cF_1 -\cF_1 \dd_A\\ 
\cF_2 (\ddH + R \nabla) + \dd_A  \cF_2 & - \cF_2\cF_1 + \dd_A^2
\end{pmatrix}~.\\[3pt]
\eeq
This can be rewritten as
\beq
\begin{split}
  \cD^2 &\= 
\begin{pmatrix}
\ddH^2 + \cR_1\cR_2 - \cF_1\cF_2 & - \ddH \cF_1 -\cF_1 \dd_A\\
\cF_2 \ddH + \dd_A \cF_2 & - \cF_2\cF_1 + \dd_A^2
\end{pmatrix}
\\[7pt]
&\qquad+
\begin{pmatrix}
\{ \ddH, (R \nabla)\}  + (R \nabla)^2 - \cR_1\cR_2 & - (R \nabla)  \cF_1\\
\cF_2  (R \nabla) & 0
\end{pmatrix}
~,
\end{split}
\eeq
where we used the definitions in \eqref{eq:Dspur2} in the above. The first term on the right hand side vanishes due to the original operator \eqref{eq:D} being nilpotent. Hence, all we need to check is the second matrix. First we check this on a $1$-form valued in $Q$, denoted $\ccY^{(1)}$, we use the same superscript on components of $\ccY$ where appropriate:
\beq\label{eq:Dsq}
\cD^2 \ccY^{(1)} \= 
\begin{pmatrix}
 \left[\{ \ddH, ( R \nabla) \}  + (R \nabla)^2 - \cR_1 \cR_2 \right] M^{(1)} - \left[(R \nabla) \cF_1\right] \aa^{(1)}\\
 \cF_2  (R \nabla) M^{(1)} 
\end{pmatrix}~.
\eeq

The first term is 
\beq\label{eq:DDeqn}
\begin{split}
 (\{ \ddH, ( R \nabla) \}  + (R \nabla)^2 - \cR_1 \cR_2 ) M^{(1)} \= &-\frac{\ap}{2} \left( (\dd_\nabla R_m{}^t{}_s) \nabla_t  M^s\right. \\[5pt]
 &\left.- R_m{}^t{}_s R_t{}^s{}_r  M^r \right) + \frac{\ap}{4}  R_m{}^t{}_s R_r{}^s{}_t M^r~,
\end{split}
\eeq
where we have truncated terms of $\cO(\ap^2)$.
The first term on the RHS of \eqref{eq:DDeqn} is
\beq\label{eq:dR}
-\frac{\ap}{2} (\dd_\nabla R_m{}^t{}_s) \nabla_t  M^s \= -\frac{\ap}{2} (\dd_\nabla R)_m{}^t{}_s \nabla_t M^s + \frac{\ap}{2} (\nabla_m R)^t{}_s \nabla_t  M^s~.
\eeq
The first term on the RHS of \eqref{eq:dR} vanishes identically; for the second term, we use that for a $3$-form $\t$
\beq
\p_{\rep{1}} (\t) \star 1 \= \t_{mnp} \phi^{mnp} \star 1 \= 6 \, \t \psi~.
\eeq
Since $\nabla_m \psi = 0$ and $R\w \psi = 0$, 
\beq
\p_{\rep{1}} \left( -\frac{\ap}{2} (\dd_\nabla R_m{}^p{}_t) \nabla_p  M^t \right) \= 0~.
\eeq
The second line of \eqref{eq:DDeqn} vanishes by use of the Bianchi identity $R_{ps}{}^t{}_q + R_{qp}{}^t{}_s + R_{sq}{}^t{}_p = 0$ and symmetries $R_{mnpq} = R_{pqmn}$, $R_{pqmn} = -R_{qpmn}$, where $R$ is the Levi-Civita curvature.\\
The second term in the first row of \eqref{eq:Dsq} vanishes as it is manifestly higher order:
$$
- \left[(R \nabla) \cF_1\right] \aa^{(1)} \= \cO(\ap^2)~.
$$
The second row is
\beq
\cF_2  (R \nabla) M^{(1)} \=  -\frac{\ap}{2} R_m{}^t{}_p (\nabla_t M^p) F^m~.
\eeq
The instanton equation implies that $\pi_7(R)=0$. It follows that $R$ is in the $\rep{14}$.  That being so, it turns out that any contraction such as $R_s R^s$ or $R_s F^s$  is also in the $\rep{14}$ -- this is shown in say equation (A.28) of \cite{delaOssa:2016ivz}. We are required to project \eqref{eq:DDeqn} to the $\rep{1}$ and so the second row of \eqref{eq:Dsq} vanishes.

We now need to check $\chD^2$ on a 0-form $\ccY^{(0)}$. The calculation precedes in much the same way. We have
\beq\label{eq:Dsq0}
\cD^2 \ccY^{(0)} \= 
\begin{pmatrix}
 \left[\{ \ddn, ( R \nabla) \}  + (R \nabla)^2 - \cR_1 \cR_2 \right] M^{(0)} - \left[(R \nabla) \cF_1\right] \aa^{(0)}\\
 \cF_2  (R \nabla) M^{(0)} 
\end{pmatrix}~.
\eeq
The first term is 
\beq\notag
\begin{split}
  \left[\{ \ddn, ( R \nabla) \}  + (R \nabla)^2 - \cR_1 \cR_2 \right] M^{(0)} &\= -\frac{\ap}{2} \dd_\nabla(R_m{}^p{}_t \nabla_p M^{(0)\,t}) - \frac{\ap}{2} R_m{}^p{}_t \nabla_p \dd_\nabla M^{(0)\,t}\\
  &\quad + \frac{\ap}{4} \tr (M^{(0)}_p R^p R_m) + \cO (\ap^2)~.
\end{split}
\eeq
The first line simplifies after using the Bianchi identity $\dd_\nabla R_m = \nabla_m R$ and the instanton condition $\psi \w R = 0$, which comes from the projection to the $\rep{7}$. We are left with 
$$
-\frac{\ap}{2} \left( R_m{}^p{}_t R_p{}^t{}_q   + \half R_q{}^p{}_t R_m{}^t{}_p\right)M^{(0)\,q} \= 0~.
$$
where the vanishing follows by use of the previously used Bianchi identity $R_{ps}{}^t{}_q + R_{qp}{}^t{}_s + R_{sq}{}^t{}_p = 0$ and symmetries $R_{mnpq} = R_{pqmn}$, $R_{pqmn} = -R_{qpmn}$ of the Levi-Civita curvature. 

The second term of first row in \eqref{eq:Dsq0} is manifestly order $\ap^2$ and so vanishes to this order. The final term is the second row in \eqref{eq:Dsq0}  is
$$
-\frac{\ap}{2}\psi \w R_m{}^p{}_t (\nabla_p M^t) F^m \= 0~,
$$
by Lemma 5 of \cite{delaOssa:2017pqy} and as $R,F$ are both instantons and so sit it in the $\rep{14}$ of $G_2$.  Hence, $\chD^2$ vanishes on a $0$-form. This means we can define a complex 
\beq
0 \rightarrow \L^0(Y,Q) \xrightarrow{\Dch}  \L^1(Y,Q) \xrightarrow{\Dch}  \L^2_{\rep{7}} (Y,Q) \xrightarrow{\Dch}  \L^3_{\rep{1}}(Y,Q) \rightarrow 0~.\label{eq:Dchcomplex}
\eeq
It should be noted that the above is only a complex modulo $\cO(\ap^2)$ corrections, that is $\Dch^2=\cO(\ap^2)$. This makes the study of the corresponding cohomology theory somewhat murky. Indeed, the mathematical framework for studying "approximate instantons" and cohomologies is quite involved, and should be avoided if possible. We will shed some more light on possible solutions to this issue here, and comment further in section \ref{sec:Double}.

To be more precise what we mean by this is we are studying the system order-by-order in $\ap$ in the spirit of effective quantum field theory. The parameter $\ap$ is a small formal parameter (with $l_s=1$ it is dimensionless) and so not only are the equations of motion, the supersymmetry varations and fields formally expanded in $\ap$ but so is this operator
$$
\Dch = \Dch_0 + \ap \Dch_1 + \ap^2  \Dch_2 + \ldots~,
$$
where $\Dch_1$, $\Dch_2$, etc, do not contain explicit factors of $\ap$. While we have determined, via the supersymmetry equations, $\Dch_0, \Dch_1$, we do not know $\Dch_2$ as this requires studying the supersymmetry equations correct to order $\ap^2$. These have not been written down yet. Furthermore, what we have shown is that
$$
(\Dch_0)^2 \= 0~,\quad \Dch_1 \Dch_0 + \Dch_0\Dch_1 \= 0~, \quad\Leftrightarrow \quad\Dch^2 = \ap^2 (\cdots)~,
$$
where the $(\cdots)$ may or may not be zero; at this point we cannot say one way or another as we cannot determine what expression to write for $\Dch_2$. The important point here is that this operator squares to zero modulo $\ap^2$ terms, which as we are working to first order in $\ap$ is good enough to do a self-consistent effective field theory analysis. Indeed, we pursue this order-by-order analysis of the operator and fields further in section \sref{sec:Double}, in which we show the operator takes an upper-triangular structure. 

Elements of $\L^1(Y,Q)$ in the kernel of $\Dch$ then parametrise the infinitesimal deformations of the heterotic $G_2$ system. There are shifts by $\Dch$-exact terms. They are generated by an action on a  $Q$-valued 0-form $\Phi$:
\beq\label{eq:DchExact}
 \cD \Phi \=
\begin{pmatrix}
\dd_\nabla + R\nabla & - \cF_1\\
\cF_2 & \dd_A
\end{pmatrix}
\begin{pmatrix}
\ve_m\\
\phi
\end{pmatrix} 
\=
\begin{pmatrix}
\dd_\nabla \ve_m - \frac{\ap}{2} R_m{}^{st}  \nabla_s \ve_t - \frac{\ap}{4} \tr (\phi F_m)\\
\ve_m F^m + \dd_A \phi
\end{pmatrix}~.
\eeq
At this point we digress briefly to consider small gauge transformations of the $G_2$ fields. The analysis of section 2 of \citeSG can be recycled here with relatively trivial changes:
\beq \label{eq:smallgauge}
\begin{split}
 \d g_{qm} &\sim \d g_{qm} + \nabla^\LC_q \ve_m + \nabla^\LC_m \ve_q~, \\
  \ccB_{qm} &\sim \ccB_{qm} + \ve^p H_{pqm} + \frac{\ap}{2} \tr \phi F_{qm}  - \ap R_{qm}{}^{ts} \nabla_t \ve_s + (\dd \mfb)_{qm}~,\\
\aa &\sim \aa + \ve^p F_p + \dd_A \phi~.
\end{split}
\eeq
Then, using $M_{qm}  = \half (\d g  + \ccB)_{qm}$ we find
\beq
M_{qm} \sim M_{qm} +   \nabla^\H_q \ve_m  + \frac{\ap}{4} \tr \phi F_{qm} -   \frac{\ap}{2} R_{qm}{}^{ts}\nabla_t \ve_s + \half \nabla^\LC_m (\ve-\mfb)_q - \half \nabla^\LC_q (\ve-\mfb)_m~.
\eeq
If we set $\mfb_m = \ve_m$ then this reproduces the first line of \eqref{eq:DchExact}. The second line follows trivially from \eqref{eq:smallgauge}. The shifts by $\Dch$-exact terms should be thought of as infinitesimal gauge transformations, provided $\mfb_m = \ve_m$, a type of initial gauge fixing. This then implies that the infinitesimal deformations of the heterotic $G_2$ system are related to $H^1_{\Dch}(Y,Q)$.\footnote{As detailed above, this cohomology is ill-defined as $\Dch$ is not exactly nilpotent. In section \ref{sec:Double} we utilise the $\ap$ expansion to define an exactly nilpotent operator and corresponding cohomology.} In fact, there is a natural choice of gauge - or representative - given by demanding the $G_2$ moduli metric we discuss later on reduce to the $SU(3)$ moduli space in an appropriate limits. We discuss this later. 

We need to gauge fix to describe the physical moduli. As we will discuss in the next section, there is a preferred gauge when reducing to $SU(3)$ in order to recover holomorphic gauge and the D-terms. The gauge fixing   is 
\beq\label{eq:gaugefix}
\nabla^{\H\, q}M_{qm} - \frac{\ap}{2} R_m{}^{spt}  (\nabla_t M_{sp})  - \frac{\ap}{4} \tr  F_{ms} \aa^s \= 0~,
\qquad \dd_A^\dag \aa + F^{qm}  M_{qm} \= 0~.
\eeq
Indeed,   $ M$ is really a deformations of the metric and $B$-field, this is a  generalisation of harmonic gauge that incorporates an $\ap$--correction. The simplest way to see this condition fixes the gauge freedom is to work order-by-order in $\ap$: expand $\ve  = \ve_0 + \ap \ve_1 + \cdots$, $\phi = \phi_0 + \ap \phi_1 + \cdots$ etc. Then the first equation in \eqref{eq:gaugefix} implies $\ve_0$ is the zero-mode of a Laplacian on a compact manifold and so vanishes. The second equation forces $\phi_0 = 0$. Returning to the first equation, we then find at order $\ap$, that $\ve_1$ is also a zero-mode of the Laplacian and so vanishes.  Hence to first order in $\ap$, the gauge freedom is fixed. If we didn't know anything about string theory, this would be a crazy gauge fixing condition to pick. Instead, we demonstrate a beautiful reason for its choice in \sref{s:adjoint}~.

In the next section, when we discuss dimensional reduction, \eqref{eq:smallgauge} will reduce to the $SU(3)$ shifts of the $\Dbar$-operator. This can also be checked by reducing the component expressions above. As discussed in \citeE in the context of $SU(3)$ manifolds, we do not really have the freedom to do $\Dbar$-shifts on the physical moduli as doing so takes us out of holomorphic gauge and violates the D-terms. Instead, the physical moduli are a harmonic representative of a bundle valued cohomology defined by the $\Dbar$-operator. This means that although we can use a cohomology to derive the algebraic structure of the moduli space, the physical deformations are a particular representative of that cohomology. Presumably there is an analogous story here, and it would be interesting to explore the cohomology of this space further in future work. 

The choice of gauge \eqref{eq:gaugefix} is natural from the point of view of reduction to $SU(3)$. When we reduce to $SU(3)$ in the next section,  it will be important that the moduli space will pick up a complex structure; we will associate holomorphic tangent vectors of the moduli space to deformations of certain fields in a natural way, and that is holomorphic gauge \citeSG. Having done so, the gauge fixing equation actually precisely recovers the D-term equations derived in \citeE (see equations (5.15-5.16) of that paper)!  The D-terms, F-terms and gauge fixing of the $SU(3)$ manifold when lifted to $G_2$ are  mixed together. We return to this later.

\section{Reduction to \texorpdfstring{$SU(3)$}{SU(3)}}
A good consistency check of this calculation is to reduce the $D$ operator to $SU(3)$. We take the $G_2$ manifold to be of the form $\Y = \X \times S^1$, where we allow the radius of the circle to decompactify and become part of $\IR^{3,1}$ for an $SU(3)$ structure compactification. Under these conditions the $G_2$ structure decomposes as
\beq\label{eq:G2toSU3}
\begin{split}
\varphi &\= \dd r \wedge \o + \Re (\O)~,\\
\psi &\= - \dd r \wedge \Im (\O) + \half \o \wedge \o~,
\end{split}
\eeq
where $r$ is the coordinate along $\IR$. We assume no field dependence on the $r$ coordinate. Given these assumptions, the $G_2$ system reduces to the Strominger-Hull system \cite{Strominger:1986uh,Hull:1986kz} and we  expect to recover the results of \citeE, \citeES obtained by removing the spurious moduli fields directly in the Strominger-Hull system.

In the reduction the instanton condition becomes
$$
F\w \psi ~~\Longrightarrow~~ F\w\o\w\o \= 0~, \qquad F\w\Psi \= 0~.\\[2pt]
$$
Acting with \eqref{eq:Dtilde} on \eqref{eq:Y}, we have
\beq\label{eq:DY}
\cD \ccY \= 
\begin{pmatrix}
\dd_\nabla   M_{m} - \frac{\ap}{2} R_m{}^{st} \nabla_s   M_t + \frac{\ap}{4} \tr \left( \aa F_m \right)  \\[5pt]
 -   M_{ m} F^m + \dd_A \aa
\end{pmatrix}~;\\[2pt]
\eeq
recall $F_m = F_{mq} \dd x^q$, $R_m{}^t{}_s = R_{qm}{}^t{}_s \dd x^q$, and $ M_s =  M_{qs} \dd x^q$. 

The co-tangent bundle $T^*Y$ reduces to the sum $T^{*(1,0)}{}\X \oplus T^{(1,0)}{}\X$, where we have used the natural isomorphism between tangent and cotangent bundles on $X$ via the hermitian metric;  we also use the natural pullback of the bundle endomorphisms from $Y$ to $X$: $\L^p(X,{\rm End}(V))$ where $V$ is now the gauge bundle on $\X$.  

The reduction ans\"atz is that   $ M_{mn} = \half \left( \d g + \ccB \right)_{mn}$ where $g$ is the $SU(3)$ metric and $\ccB = \d y^a \ccB_a$ is a gauge invariant deformation of the $B$-field with respect to gauge transformations of $B$, see e.g. \citeES. Notice we are doing this with respect to the redefined field. We find, using \eqref{eq:G2toSU3}, that
\beq \label{eq:MtoX}
\begin{split}
   M_{\rb \mb} &\= \d y^\a \left( \D_{\a\, \rb \mb} + \half \ccZ_{\a\, \rb \mb}\right) = \d y^\a \D_{\a\,\rb \mb}~,\\
  M_{\sb \m} &\= \half \d y^\a \ccZ_{\a\,\sb \m} ~,
\end{split}
\eeq
where in the first line we have taken the $SU(3)$ manifold to be in holomorphic gauge \citeSG in which $\ccZ^{(0,2)}  = 0$.  We have evaluated the field components in the complex structure $J$ of $X$. We are in the fortunate position that our $G_2$ field conventions reduce naturally to the $SU(3)$ conventions in \citeSG. The key ones are
\beq\label{eq:holgauge}
\ccZ_\a \= \ccB_\a  + \ii \del_\a \o~, \quad \ccZb_\a \= \ccB_\a - \ii \del_\a \o~, \quad \d A^{0,1} = \d y^\a \fD_\a \A \= \aa~,
\eeq
where $\a$ denotes a deformation with respect to a holomorphic parameter $ y^\a$ -- the moduli space becomes complex  \K upon reduction with $d=4,N=1$ supersymmetry -- and $\del_\a \o$ is a holomorphic deformation of the hermitian form $\o$  on $X$. A holomorphic deformation of complex structure $J$ on $X$ is denoted $\D_{\a\,\nb}{}^\m \dd x^\nb$ and is $\delb$-closed. The field $\D_{\a\,\mb\nb}$ comes from using the metric to lower an index. In the following we will suppress the $\d y^\a$ and subscript $\a$, understanding all deformations to be holomorphic, unless there is possible ambiguity in which case we write in all detail. We also take $\ccZ$ to be complex type $(1,1)$ and so do not write this explicitly. \footnote{The $(0,2)$-component of $\ccZ$ is present as the antisymmetric part of  $\D_{\mb\nb}$.}  Similarly, a holomorphic deformation of $A^{(0,1)} = \A$ is denoted $\aa$.

Putting this ansatz into \eqref{eq:DY} and using \eqref{eq:G2toSU3},  we project on to the $(0,2)$-component of $\check \cD \ccY$. The first row of \eqref{eq:DY} gives two components as the real index of $Y$ decomposes as $m=(\m,\mb)$ upon reduction to $X$, while the second row decomposes naturally:
\beq
\begin{split}
( \check \cD  \ccY)^{(0,2)}{}_\m &\= - \half \delb \Z_\m - \ii (\del \o)_{\r \m} \D^\r - \frac{\ap}{2} R_\m{}^\r{}_\s \nabla_\r \D^\s - \frac{\ap}{4} R_{\m}{}^\s{}_{\r} \nabla^\r \Z_\s + \frac{\ap}{4} (\a F_\m)~,\\[4pt]
( \check \cD  \ccY)^{(0,2)}{}^\n &\=  \delb \D^\n ~,\\[4pt]
(\check \cD  \ccY)^{(0,2)}{}^A{}_B &\= \left(- \D^\r F_\r{} + \delb_\A \aa\right)^A{}_B~,\\[4pt]
\end{split}
\eeq
where $\ccZ_\m = \ccZ_{\m\nb} \dd x^\nb$, $R_m{}^n{}_p = R_{qm}{}^n{}_p \dd x^q$ and $F_m = F_{mn} \dd x^n$. We have used capital letters to denote the representation $\End(V)$. The projection onto the $\rep{7}$ under reduction amounts to taking the trace with the hermitian metric and wedging with the holomorphic volume form of $X$ both of which are automatically satisfied.

This can be equivalently written as
\beq\label{eq:moduli3}
(\check \cD \ccY)^{0,2} \=
\begin{pmatrix}
\half \delb \Z - \ii (\del \o)_\r \D^\r + \frac{\ap}{2} R^\r{}_\s \nabla_\r \D^\s + \frac{\ap}{4} R^\r{}_\s \nabla^\s \Z_\r - \frac{\ap}{4} (\a F) \\[7pt]
\left( \delb \D^\m \right) g_{\m \nb} \, \dd x^\nb\\[7pt]
- \D^\r F_\r + \delb_\A \aa
\end{pmatrix}~.
\eeq
At this point we compare with the $\Dbar$ operator constructed directly in the $SU(3)$ construction after eliminating the spurious modes \cite{McOrist:2021dnd,McOrist:2024zdz}. The immediate observation from \citeE is that the moduli equations are not quite the same. The resolution is simple however, we need to perform a  field redefinition
\beq\label{eq:Zrdef}
 \ccZ^{(1,1)}\=\wt  \ccZ^{(1,1)}+ \frac{\ap}{2} \dd x^\r \nabla^\m \nabla_\r  \wt \ccZ_{\m}^{(0,1)}~,
\eeq
and following the analysis in section 5 of \citeE, the first line of \eqref{eq:moduli3} can be rewritten as
\beq\label{eq:moduli3b}
\half \delb \wt \ccZ  -\ii \,   \D{}^\r (\del\o)_\r + \frac{\ap}{2}  R^\r{}_\s\nabla_\r \D{}^\s  -\frac{\ap}{4} \tr \big( \aa\, F \big)  ~.
 \eeq
This is now precisely the $\Dbar$-operator without the spurious modes constructed in \citeE. 

We can also consider the $(1,0)$-component of $\cD$ acting on  $\ccY^{0,1}$. Part of this will be the action of the holomorphic $D$-operator of \citeES. We get\\[3pt]
\beq\label{eq:holDY}
\begin{split}
\!\cD^{(1,0)} \ccY^{(0,1)}{}_\m &\= -\del \Z_\m + \dd x^\n \G^\H{}_\n{}^\r{}_\m \Z_\r \= -g_{\m \sb} \, \del_{\nabla^\CH} \Z^\sb~,\\[5pt]
\!\cD^{(1,0)} \ccY^{(0,1)}{}_\mb &\= 2\del \D_{\r \mb} \dd x^\r +H_{\rb \mb} \Z^\rb  -\ap R_{\mb}{}^\b{}_\g \nabla_\b \D^\g + \frac{\ap}{2} R_{\mb}{}^\bb{}_\gb \nabla_\bb \Z^\gb + \frac{\ap}{2} \tr \big( \aa F_\mb) \big)~,\\[5pt]
\cD^{(1,0)} \ccY^{(0,1)}{}^A{}_B  &\= \left( \Z_\m F^\m + 2\del_{\A^\dagger} \aa\right)^A{}_B~.
\end{split}
\eeq
After applying the field redefinition \eqref{eq:Zrdef} we find equation (3.8) of \citeES for a $D$-operator. Together with the $\Dbar$-operator these define a real operator that satisfies a Hermitian Yang-Mills equation. 

Now we use the information that $\check \cD \ccY = 0$ describes the moduli. The reduction to the $(0,2)$-component of this equation gives precisely the kernel of the $\Dbar$-operator after the field redefinition.

Furthermore, as $\Dch^2 = 0$ we find a shift by $\Dch$-exact terms \eqref{eq:DchExact}: $\ccY \to \ccY + \Dch \Phi$ which has an interpretation in the language of the $SU(3)$ case:
\beq\label{eq:SU3gauge}
\begin{split}
\wt \cD \Phi^{(0,1)}{}_\m &\= \delb \ve_\m - \ii \ve^\r (\del \o)_{\r \m} - \frac{\ap}{2} R_\m{}^\r{}_\s \nabla_\r \ve^\s - \frac{\ap}{2} R_\m{}^\rb{}_\sb \nabla_\rb \ve^\sb - \frac{\ap}{4} \tr (\phi F_\m)~,\\[5pt]
\wt \cD \Phi^{(0,1)}{}^A{}_B &\= (\ve^\r F_\r + \delb_\A \phi)^A{}_B~,\\[5pt]
\wt \cD \Phi^{(0,1)}{}_\nb &\= g_{\r \nb} \, \delb \ve^\r~.
\end{split}
\eeq
These are to be compared to the discussion of $\Dbar-$exact shifts of the operator constructed in \citeE, e.g. section 4 of that paper. The transformations \eqref{eq:SU3gauge} are a subset of small gauge transformations of the $SU(3)$ problem -- $\ccZ^{1,1}$ does not shift by a $\del$-exact term -- however, as noted in \citeE, these transformations are not a gauge symmetry of the physical solution.  The gauge transformations are real and the $\Dbar$-operator is constructed in holomorphic gauge, and holomorphic gauge completely fixes  gauge transformations.   Said differently, the  $SU(3)$ heterotic vacua corresponds to the intersection of $\Dbar\ccY = 0$ (F-term)  and its adjoint $\Dbar^\dag \ccY = 0$ (D-term). The shift by a $\Dbar$-exact term preserves the F-term but not the D-term condition. The solution is the $\Dbar$--harmonic representative of the bundle valued cohomology group constructed in \citeE. 
We will investigate this further in the context of $G_2$ below.

Finally, in \citeES it was shown that $D+\Dbar$ for the $SU(3)$ structure case satisfied a hermitian Yang-Mills equation provided the supersymmetry and Bianchi identity held.

\section{Moduli space metric and an adjoint for \texorpdfstring{$\Dch$}{Dcheck}}
\label{s:adjoint}
\subsection{A proposal for the  \texorpdfstring{$G_2$}{G2} heterotic  moduli space metric}
We propose a $G_2$ moduli space metric, correct to first order in $\ap$. It is given by\footnote{It is interesting to note that we could define a field $\wt M_{mn} = M_{mn} - \frac{\ap}{4} R_{mpnq} M^{pq}$ in which the metric becomes $\int \wt M_{mn} \wt M^{mn} + \frac{\ap}{4} \tr \aa_q \aa^{q\,\dag}$, which is a Weyl-Peterson inner product. However, all is not for free: the field $\wt M$ does not transform like $\d g + \ccB$, instead it has a transformation law complicated by the $\ap$-term.  
}

\beq\label{eq:modulimetric2}
\langle  M,  M \rangle   \= \frac{1}{2} \int_Y \vol \, \left(  M_{mn}    M^{mn}  +\frac{\ap}{2}  M^{mn}  M^{pq} R_{pmqn} + \frac{\ap}{4} \tr \aa_q \aa^{q\,\dag}  \right) + \cO(\ap^2)~.
\eeq
At this point it is instructive to reduce   the gauge fixing condition \eqref{eq:gaugefix} to $SU(3)$ using the ansatz \eqref{eq:MtoX}--\eqref{eq:holgauge} in the previous section. We allow only holomorphic variations of parameters, so that only $\D_{\mb\nb}$ and $\ccZ_{\m\nb}$ are non-zero\footnote{We have adopted the notation that $\D^\m \cong \d y^\a \D_\a{}^\m$ and $\bar \D^\nb \cong \d y^\bb \D_\bb{}^\nb$ for the holomorphic and antiholomorphic variations of complex structure. Similarly, $\ccZ_{\m\nb} \cong \d y^\a \ccZ_{\a\, \m\nb}$ and $\ccZb_{\m\nb} \cong \d y^\bb \ccZb_{\bb\m\nb}$. }. This is holomorphic gauge in \citeSG. After doing some work \eqref{eq:gaugefix} becomes
 \beq\label{eq:eq}
\begin{split}
& \nabla^{\Bi \,\mb} \D_{\mb}{}^\n - \half H^{\s\tb\n} \ccZ_{\s\tb}  - \frac{\ap}{4} \tr F^\n{}^\lb \aa_\lb + \frac{\ap}{4} R^\n{}^{\mb\s\tb} \nabla_\mb \ccZ_{\s\tb} +\frac{\ap}{2} R^\n{}^{\sb\t\mb} \nabla_\t \D_{\sb\mb} \= 0~,\\[5pt]
&\delb^\dag \ccZ_\n \= 0~,\\[5pt]
&\delb_\A^\dag \aa  +  \half F^{\r\nb} \ccZ_{\r\nb} \= 0~.
\end{split}
 \eeq
 In doing this calculation we have used that $\nabla^{\B \,m}  M_{mn} = \nabla^{\H\,m}  M_{mn} +H^{pq}{}_n  M_{pq}$, with the dilatino variation. In the first line of \eqref{eq:eq} we have used the Bismut connection; in contrast the Hull connection is used for the same term in \citeE. While the difference is order $\ap^2$ the Bismut connection appears more suitable  to studying higher order corrections in $\ap$. 
 
We see that what appeared as a gauge fixing in the $G_2$ compactification becomes, under a suitable choice of holomorphic coordinates, the D-terms in the $SU(3)$ reduction. The D-terms and gauge fixing are intertwined. There is in fact a beautiful explanation for this which we come to shortly. 

Returning now to  \eqref{eq:modulimetric2} the second term is familiar: it looks like the unification of the $\ap$-correction involving the Riemann curvature to the moduli space metric computed in holomorphic gauge in \citeSG. This strongly suggests dimensionally reducing the moduli space metric which we now do. 

Using \eqref{eq:MtoX} we find \eqref{eq:modulimetric2} becomes\\[5pt]
\beq\label{eq:DimRed1}
\!\!\langle  M,  M \rangle|_{SU(3)}   = \int_X\left[ \D^{\m} \star \bar \D^\nb g_{\m\nb} + \frac{1}{4} \ccZ \star \ccZb  + \frac{\ap}{4} \tr \aa \star  \aa  + \frac{\ap}{2}\left( \D_{\nb\sb} \,\bar\D_{\r\m} +\frac{1}{4} \ccZ_{\m\nb} \,\ccZb_{\r\sb} \right) R^{\nb\r\sb\m} \vol \right].\\[5pt]
\eeq
We see we get precisely the  $SU(3)$) moduli space metric correct to first order in $\ap$ without the spurious degrees of freedom written in \citeSG. It would be a good check that this is the same metric that comes from a  dimensional reduction of the action \eqref{eq:10daction} over a $G_2$ manifold following the prescription laid out in \citeM. Setting that aside, we will now give a non-trivial consistency check for this moduli space metric by constructing an adjoint operator $\Dch^\dag$. 

\subsection{Adjoint operator for \texorpdfstring{$\Dch$}{Dcheck}}
Using the moduli space metric \eqref{eq:modulimetric2} we can now construct an adjoint operator $\Dch^\dag$ following the logic laid out in \citeE. That is, we extend \eqref{eq:modulimetric2} to be an inner product on $d$-forms valued in $Q$ in an obvious way:
\beq\label{eq:modulimetricG2b}
\langle \ccY^d, \ccY^d \rangle \= \frac{1}{2V} \int_Y   \left(   M_{m}    \star  M^{m}  + \frac{\ap}{2}(  M^{mn} \star   M^{pq}) R_{pmqn}  + \frac{\ap}{4} \tr \aa\star \aa \right)~,
\eeq
where $M_m  = M_{Pm} \dd x^P$, with $P$ the multi-index for the $d$-form; $M_{mn} = M_{P' mn} \dd x^{P'}$ with $P'$ the $d-1$ form indices, and the middle term is taken to be zero if $d=0$; $\aa$ is understood to be a $d$-form with form indices suppressed. 

Then as for \eqref{eq:DchExact} we let $\Phi$ be a $0$-form valued in $Q$. We define the action of the adjoint on a $1$-form $\ccY$ --  the form of interest to the moduli problem -- as
\beq
\langle \ccY, \Dch \Phi \rangle \= \langle \Dch^\dag \ccY, \Phi \rangle~.
\eeq
We can calculate what this is in components
\beq
\begin{split}
 \langle \ccY, \Dch \Phi \rangle &\= \frac{1}{2V} \int_Y \left\{ M^{qm} \left( \nabla_q^\H \ve_m - \frac{\ap}{2} R_{qm}{}^s{}_t \nabla_s \ve^t + \frac{\ap}{4} \tr (\phi F_{qm}) \right) \right. \\
 &\quad \left. +\frac{\ap}{2}  M^{qm} (\nabla^s \ve^t) R_{sqtm} + \frac{\ap}{4} \tr \left[( \ve^t F_t{}_q + \nabla_{A\,q} \phi) \aa^q\right] \right\}~.
\end{split}
\eeq
Using the Bianchi identity for the Riemann curvature for Levi-Civita, $R_{qms}{}^t + R_{qsm}{}^t = - R_{smq}{}^t$, integrating by parts and the dilatino equation \eqref{eq:10dsusy}, we find
\beq
\begin{split}
  \langle \Dch^\dag \ccY,  \Phi \rangle &\= \frac{1}{2V} \int_Y \left\{ \ve_t \left(-\nabla_q^\H M^{qt} +\frac{\ap}{2} \nabla^s M^{qm} R_{smq}{}^t +  \frac{\ap}{4} \tr (F^{tq} \aa_q) \right)  \right.\\
  &\qquad +\frac{\ap}{4} \tr \phi \left(  \dd_A^\dag \aa + M^{qm} F_{qm}\right)~.
\end{split}
\eeq
From this we can read off the adjoint in matrix notation
\beq
\Dch^\dag \ccY \= 
\begin{pmatrix}
 -\nabla^{\H\, q}M_{qm} + \frac{\ap}{2} R_m{}^{spt}  (\nabla_t M_{sp})  + \frac{\ap}{4} \tr  F_{ms} \aa^s \\
 \dd_A^\dag \aa + M^{qm} F_{qm}
\end{pmatrix}~.
\eeq
We have written the action of the adjoint on a $1$-form as that is the physically relevant case. The extension to a $p$-form is straightforward given the metric \eqref{eq:modulimetricG2b} and an appropriate projection to $G_2$-representation compatible with the complex \eqref{eq:Dchcomplex}.
 
We are now in the position to do a very nice consistency check that the moduli space metric we have proposed in  is correct: the gauge fixing condition \eqref{eq:gaugefix} is precisely 
$$
\Dch^\dag \ccY \= 0~.
$$ 
Hence, the physical moduli of the $G_2$ compactification are described by 
\beq\label{eq:moduliEqn}
\Dch \ccY \=0 ~, \qquad \Dch^\dag \ccY \= 0~.
\eeq
 This looks formally very similar to the F-terms and D-terms written down in \citeE, though of course there is no such thing in a $G_2$ compactification as the spacetime supersymmetry is $d=3$ and $N=1$. Instead, the first equation implies $\ccY$ is an element of a certain cohomology; the second equation corresponds to gauge fixing and picks out the harmonic representative of that cohomology. This is similar to the conclusion in \citeE in which the $SU(3)$ moduli were described as the harmonic representative of a certain cohomology. 
 
 As promised the underlying reason why \eqref{eq:gaugefix} reduces to the D-terms provided we fix holomorphic gauge. It is simply because they can be written as $\Dch^\dag \ccY=0$ and it is this equation which naturally reduces to the D-term equation in $SU(3)$ (viz. $\Dbar^\dag \ccY = 0$ in the notation of \citeE).  While this was initially a surprise,  especially as \eqref{eq:gaugefix} starts life without knowledge of the moduli space metric, the existence of the moduli space metric allows us to write down  \eqref{eq:moduliEqn}, and then the fact we get the $SU(3)$ D-terms and F-terms upon reduction is  natural. 
 
\section{A double extension for \texorpdfstring{$G_2$}{G2} moduli}
\label{sec:Double}
Our goal here is to rewrite the heterotic $G_2$ differential and its cohomology in a way so that it is triangular. Recall, to be in the kernel of $\check D$ to the appropriate order in $\ap$, elements $\ccY\in\L^{d}(Y, Q)$, where 
\beq\
\ccY \= 
\begin{pmatrix}
 M \\
 \aa
\end{pmatrix}~,
\eeq
must satisfy
\beq
\chD\ccY\= \pi\circ\wt\cD\ccY\=0+\cO(\ap^2)\:,
\eeq
where $\pi$ denotes the projection onto the appropriate irreducible $G_2$ representation. Note that this equation is equivalent to
\beq
\Psi\wedge\wt\cD\ccY\=0+\cO(\ap^2)\:,
\eeq
where $\Psi=e^{-2\phi}\psi$ is the re-scaled $\dd$-closed four-form.

Let us rephrase the heterotic $G_2$ moduli complex in a slightly more convenient way for the current section:
\beq
\label{eq:ModuliComp2}
0\rightarrow\L^{0}(Y, Q)\xrightarrow{\wt\cD}\L^{1}(Y, Q)\xrightarrow{\Psi\wedge\wt\cD}\L^{6}(Y, Q)\xrightarrow{\wt\cD}\L^{7}(Y, Q)\rightarrow0\:.
\eeq
By the computations of section \ref{sec:DiffComp}, it is straight-forward to check that this is also a differential complex modulo $\cO(\a^2)$ corrections. Furthermore, it is equivalent to the heterotic moduli complex \eqref{eq:Dchcomplex}. Next, we expand the $G_2$-structure $\Psi$ together with our fields $\ccY$ in $\ap$. That is, we write
\begin{align}\notag
 M&\=  M_0 + \ap  M_1 + \cO(\ap^2)~,\\[5pt]
\Psi&\=\Psi_0+\ap\Psi_1+\cO(\ap^2)\:.\notag
\end{align}
It is implicit that $\aa$ is already first order in $\ap$ -- the Yang-Mills equations come normalised with a factor of $\ap$ -- and so we do not need to expand $\aa$ in $\ap$. Note also that $\Psi_0=\psi_0$, the $G_2$ four-form of the $G_2$ holonomy background. 

For $\ccY\in\L^{0}(Y, Q)$ and $\ccY\in\L^{6}(Y, Q)$ we may then consider
\beq\notag
{\footnotesize
\wt\cD\ccY=
\begin{pmatrix}
 \dd_\nabla  M_{0m} +\dd_\nabla(\ap  M_{1m})-\ap {T^n}_m M_{0n} - \frac{\ap}{2} R_m{}^p{}_t \nabla_p  M_0^t - \frac{\ap}{4} \tr \left( F_m\aa \right)\\
 F^m M_{0m} + \dd_A \aa
\end{pmatrix}+\cO(\ap^2)~,}
\eeq
while for $\ccY\in\L^{1}(Y, Q)$ we get
\beq\notag
{\footnotesize
\Psi\wedge\wt\cD\ccY=
\begin{pmatrix}
\ap\Psi_1\wedge\dd_\nabla M_{0m}+\Psi_0\wedge\Big(\dd_\nabla  M_{0m}+\dd_\nabla(\ap  M_{1m})-\ap {T^n}_m M_{0n} - \frac{\ap}{2} R_m{}^p{}_t \nabla_p  M_0^t - \frac{\ap}{4} \tr \left( F_m \aa \right)\Big)\\
\Psi_0\wedge\left(F^m M_{0m} + \dd_A \aa\right)
\end{pmatrix}+\cO(\ap^2)~,}
\eeq
where now $\nabla$ denotes the Levi-Civita connection of the background $G_2$-holonomy manifold, $R$ is the curvature of this connection, and $\ap T$ includes the flux of the Hull-connection, and any ${\cal O}(\ap)$ corrections to the Levi-Civita connection from $\ap$ corrections to the metric. Note that these corrections to the metric, or alternatively $\Psi_1$, are determined by solving the heterotic $G_2$ system order by order in $\ap$. We also take $F$ to be an instanton with respect to the zeroth order $G_2$ structure $\Psi_0$. 

Let us then define a new operator $\wt\cD'$ which decomposes the action of the differential in the complex \eqref{eq:ModuliComp2} into forms valued in $\wt Q'= T^*Y \oplus {\rm End}(V)\oplus T^*Y$. That is, we now write
\beq\
\ccY \= 
\begin{pmatrix}
\alpha' M_1 \\
 \aa \\
 M_0
\end{pmatrix}~.
\eeq
The operator $\cD'$ is constructed so as to only contain explicit appearances of $\ap$ up to first order, and takes the following form on $\ccY\in\L^{0}(Y, Q')$ and $\ccY\in\L^{6}(Y, Q')$,
\beq\notag
\wt\cD'\ccY\=
{\footnotesize
\begin{pmatrix}
\dd_\nabla(\ap M_{1m})-\ap {T^n}_mM_{0n} - \frac{\ap}{2} R_m{}^p{}_t \nabla_p  M_0^t - \frac{\ap}{4} \tr \left( F_m \aa \right)\\
F^m M_{0m} + \dd_A \aa \\
\dd_\nabla  M_{0m}
\end{pmatrix}}\:,
\eeq
while on $\ccY\in\L^{1}(Y, Q')$ it reads
\beq\notag
\wt\cD'\ccY=
{\footnotesize
\begin{pmatrix}
\Psi_0\wedge\Big(\dd_\nabla(\ap  M_{1m}) - \frac{\ap}{4} \tr \left( F_m \aa \right) - \ap {T^n}_m M_{0n}  - \frac{\ap}{2} R_m{}^p{}_t \nabla_p  M_0^t \Big)+\ap\Psi_1\wedge\dd_\nabla  M_{0m}\\
\Psi_0\wedge\left( \dd_A \aa  + F^m M_{0m} \right) \\
\Psi_0\wedge\dd_\nabla  M_{0m}
\end{pmatrix}}\:,
\eeq
or in terms of a matrix
\beq
\wt\cD'\ccY\=
\begin{pmatrix}
\Psi_0\wedge \dd_\nabla & 
-\Psi_0\w \cF_1  &
\Psi_0\w \left( -\ap T_m + (R\nabla) \right)
+\ap\Psi_1\wedge\dd_\nabla \\
 0& \Psi_0\w \dd_A& \Psi_0\w \cF_2 \\
0&0& \Psi_0\wedge\dd_\nabla
\end{pmatrix}
\begin{pmatrix}
\alpha' M_1 \\
 \aa \\
 M_0
\end{pmatrix}~,
\notag
\eeq
where we use \eqref{eq:defnRnabcF} and $T_m( M_0) = T^n{}_m  M_{0\,n}$. The operator $\wt \cD'$ is upper triangular.
In this new decomposition, the complex \eqref{eq:ModuliComp2} may then be written as
\beq
\label{eq:RedefComp}
0\rightarrow\L^{0}(Y, Q')\xrightarrow{\wt\cD'}\L^{1}(Y, Q')\xrightarrow{\wt\cD'}\L^{6}(Y, Q')\xrightarrow{\wt\cD'}\L^{7}(Y, Q')\rightarrow0\:.
\eeq
Furthermore, within the framework of working order by order in the $\ap$ expansion, the differential $\wt\cD'$ must square to zero {\it exactly}! This is because we solve our equations exactly, at each order in the $\ap$ expansion. 

Explicitly, we have defined the operator $\wt\cD'$ so it can be decomposed into the form
\begin{equation}
\wt\cD'=\wt\cD'_0+\ap\wt\cD'_1\:,
\end{equation}
where $\wt\cD'_0$ and $\wt\cD'_1$ do not have explicit appearances of $\ap$. By a similar argument to the one found in section \ref{sec:chDNil} for $\Dch$, we have 
\begin{equation}
\wt\cD'^2={\cal O}(\ap^2)\quad\Leftrightarrow\quad\wt\cD'^2_0=0\:,\quad\wt\cD'_0\wt\cD'_1+\wt\cD'_1\wt\cD'_0=0\:.
\end{equation}
However,
\begin{equation}
\cD'^2_0=\wt\cD'^2_0+\ap(\wt\cD'_0\wt\cD'_1+\wt\cD'_1\wt\cD'_0)+\ap^2\wt\cD'^2_1\:,
\end{equation}
and $\wt\cD'_1$ is upper-triangular with zero on the diagonal. It follows that $\wt\cD'^2_1=0$, and the operator $\wt\cD'$ is nilpotent.

Decomposing the heterotic $G_2$ moduli story in this way, using the $\ap$ expansion, has two advantages. Firstly, we get a differential which squares to zero exactly, and therefore also a differential complex. This means that the corresponding cohomologies are well defined mathematically. In particular, the cohomology $H^1_{\wt\cD'}(Y,Q')$ is well-defined, and computes the infinitesimal deformations of the heterotic $G_2$ system modulo $\cO(\ap^2)$ corrections. The second advantage is that $\wt\cD'$ actually defines a double extension of complexes, just like in the $SU(3)$ case! We may therefore use homological algebra techniques to compute $H^1_{\wt\cD'}(Y,Q')$ in terms of more familiar cohomology groups. We will schematically outline how this goes, but we shall return to this in more detail in future work. 

To give some details, note that $\wt\cD'$ acts on our field vector $\ccY$ in an upper-triangular fashion. This is the hallmark of extension sequences, and it allows us to build $\L^{p}_{\wt\cD'}(Y, Q')$ as a double extension, starting with an extension of $T^*Y$ by $\End_0(V)$
\beq\notag
0\rightarrow\L^p(Y,\End_0(V))\rightarrow\L^p(Y,Q'_1)\rightarrow\L_0^p(Y,T^*Y)\rightarrow0~,
\eeq
where each complex in this sequence is of the form \eqref{eq:RedefComp}, i.e. $p\in\{0,1,6,7\}$. Here $\L_0^p(Y,T^*Y)$ is the complex governing zeroth order $\ap$ deformations of the $G_2$ holonomy background together with zeroth order deformations of the Kalb-Ramond $B$-field, while $\L^p(Y,\End_0(V))$ is the complex governing deformations of the instanton connection. This is the $G_2$ version of the Atiyah complex \cite{Atiyah:1955}, studied in the heterotic context in \cite{delaOssa:2016ivz}. The operator $\wt\cD'$ then defines $\L^{p}_{\wt\cD'}(Y, Q')$ as an additional extension of $Q'_1$ by another copy of $\L_1^p(Y,T^*Y)$, the first order $\ap$ deformations of the geometry and $B$-field
\beq
0\rightarrow\L_1^p(Y,T^*Y)\rightarrow\L^p(Y,Q')\rightarrow\L^p(Y,Q_1')\rightarrow0\:.
\eeq
Each short exact sequence lead to long exact sequences of cohomologies, which allows one to rephrase the cohomologies $H^p(Y,Q')$ in terms of more ordinary cohomologies of the individual bundles and kernels of Atiyah like maps.

\section{Conclusions}
The issue of the connection on the tangent bundle has received attention throughout the sands of time in the study of heterotic moduli at the level of supergravity. While most attention focusses on the $SU(3)$ case, it is also present at the level of $G_2$ compactifications of string theory. In both cases it is mathematically more elegant, at least with our current level of knowledge, to study the moduli of a theory where the connection on the tangent bundle is an instanton and allowed to vary independently on the moduli of the underlying compact manifold. This gives the problem a well-defined setting with many mathematical tools at hand. In contrast, the physics has a preferred connection. This is the Hull connection. It is determined in terms of the other moduli and has been checked against string scattering amplitudes. Our goal here was to include this dependence in the phrasing of the moduli space as the kernel of an operator on an auxiliary bundle $Q$. 

What is nice about the $G_2$ setting in comparison to the $SU(3)$ case, is that the D-terms and F-terms are incorporated into the vanishing of a single operator, which we have identified. We have checked in detail that when the $G_2$ operator is dimensionally reduced to the $SU(3)$ manifold one recovers the $\Dbar$, $D$ and $\Dbar^\dag$--operators for heterotic $SU(3)$ compactifications \citeES. We proposed a moduli metric for $G_2$ compactifications and checked it reduces to the $SU(3)$ metric. The $G_2$ moduli space metric allows us to define a $\Dch^\dag$, whose kernel defines a gauge fixing for the $G_2$ moduli. Upon reduction it also reduces to the D-terms in $SU(3)$ holomorphic gauge for an $SU(3)$ manifold.   Finally, we observe that in the $G_2$ setting, although the $\Dch$-operator is not upper triangular it can be rewritten in an upper triangular way by treating the problem order-by-order in $\ap$. This could open up doors to new calculations of the cohomologies underpinning the moduli. 

There are many further questions. One could try to compute the cohomology in explicit examples, both locally \cite{Gunaydin:1995ku}, and for compact geometries \cite{delaOssa:2021qlt, Lotay:2021eog, Galdeano:2024fsc}, as has been done in the $SU(3)$ case \cite{Chisamanga:2024xbm, deLazari:2024zkg}. In this context, it would be interesting to extend this calculation to include flux $H=\cO(1)$ with $AdS_3$ vacua. From the string theory point of view these are easier to define in terms of the $\ap$ expansion. This is of clear interest to problems related to the $AdS_3/CFT_2$ correspondence. This will also be instructive in the study of higher order $\ap$ corrections. A related question is what happens to higher order in deformation theory. This will involve an extension of the work in \citeS to higher order in deformation theory and finding a Maurer-Cartan equation for the moduli. In the $SU(3)$ case, this involved an $L_\infty$ algebra that truncated to $L_3$. Presumably there is an $L_\infty$ algebra underpinning the $G_2$ case: does it truncate, and if so to what order? It is not obvious what will happen as the supersymmetry preserved is half that of the $SU(3)$ case. It would also be interesting to extend the operator here to a $G_2$ universal geometry, as studied in the $SU(3)$ case in \citeUG. Again, the reduced supersymmetry makes studying this a qualitatively different question to the $SU(3)$ case.

\subsection*{Acknowledgements}
We would like to thank Javier Jose Murgas Ibarra, Magdalena Larfors, Matthew Magill, Enrico Marchetto, Sebastien Picard and Xenia de la Ossa for helpful and enlightening conversations. JM is in part supported by an ARC Discovery Project Grant DP240101409. JM, ES and MS would like to thank the MATRIX Research Institute where part of this work was completed.  
\appendix

\newpage 
\section{Some results for \texorpdfstring{$G_2$}{G2}}
\subsection*{Projections of forms into irreducible representations}
\label{s:projections}
Here we give the explicit recipe \cite{delaOssa:2017pqy} for the decomposition of forms into irreducible representations. For $0$- and $1$-forms there is nothing to do. A $2$-form $\b$ can be decomposed 
$$
\b \= \a \lrcorner \phi + \g~,  \qquad \a\in \L^1(Y)~, \qquad \g\lrcorner \phi \cong \g \w \psi \=0~,
$$
where $\a\in\L^1(Y)$ and $\g\in\L^2_{\rep{14}}$ such that $\g\w\psi=0$. We denote the contraction between a $k$-form $A_k$ and a $k+p$-form $B_{k+p}$:
\beq
A_k \lrcorner B_{k+p} \= \frac{1}{k!p!} A^{m_1\cdots m_k} B_{m_1\cdots m_kn_1\cdots  n_p} \dd x^{n_1\cdots n_p}~.
\eeq
The projections on $\b$ are then 
\beq
\begin{split}
 \pi_{\rep{7}} &\= \frac{1}{3} (\b+\b\lrcorner\psi)~,\qquad
\pi_{\rep{14}} \= \frac{1}{3} (2\b-\b\lrcorner\psi)~. \\
\end{split}
\eeq
We can write
\beq
\begin{split}
 \L^2_{\rep{7}} &\= \left\{\b\in \L^2(Y) ~\big|~   \b \lrcorner\psi \= 2\b\right\}~,\qquad
 \L^2_{\rep{14}} \= \left\{\b\in \L^2(Y) ~\big|~   \b \w \psi \= 0\right\}~.\\
\end{split}
\eeq
Any 3-form $\l$ can be decomposed as
$$
\l \= f\phi + \alpha \lrcorner \psi + \chi~,
$$
for a function $f$, a 1-form $\a\in\L^1(Y)$ and some 3-form $\chi\in\L_{\rep{27}}^3(Y)$ which satisfies
\beq
\chi\lrcorner \phi \= 0~, \qquad \chi \lrcorner \psi \=0~.
\eeq
The three-form can also be characterised in terms of a one-form valued in $TY$:
\beq
\l \= \half M^n\w \phi_{npq} \dd x^{pq}~,\qquad M\in \L^1(TY)~.
\eeq
We can form a matrix $M_{mn} =  M_m{}^p g_{pn}$. It can then be shown that $\pi_{\rep{1}}(M)$ corresponds to  $\tr M$; $\pi_{\rep{7}}(\l)$ to $\pi_{\rep{7}}(m) = -\alpha \lrcorner \phi / 3$ with $m$ the antisymmetric part of the matrix $M$; and elements in $\L^3_{\rep{27}}$ to the traceless tensor $h_{ab} = M_{(ab)} - \frac{1}{7} g_{ab} \tr M$. 

\newpage
\section{Conventions in the literature}
\label{app:literature}
We compare some conventions and results for $SU(3)$ papers. 
Here AQS refers to \cite{Anguelova:2010ed} (with same conventions used in say \cite{Becker:2007zj}, \citeE,\citeES). OS refers to \cite{delaOssa:2014cia}. MarSpar to \cite{Martelli:2010jx}. 

\subsection*{Connecting bilinears and tensors}
Suppose that 
\beq
\nabla_m \ve   \= \del_m \ve + \frac{1}{2} \Th_{mab} \S^{ab} \ve \= 0~.
\eeq

We compute the action of $\nabla$ on a vector written as a spinor bilinear, assuming $\ve$ is Grassmann even:
$$
v^m \= \ve^\dag \g^m \ve~.
$$
We have,
\beq	
\begin{split}
 \del_m v^n &\= (\del_m \ve^\dag)\g^n \ve + \ve^\dag \g^c \, \del_m E_c^n\, \ve + \ve^\dag \g^n \del_m \ve  \\
&\= (\nabla_m \ve^\dag)\g^n \ve + \ve^\dag \g^n \nabla_m \ve + \frac{1}{4} \Th_{mab} \ve^\dag[\g^{ab} ,\g^c] \ve \, E_c^n  + v^c \del_mE_c^n\\
&\=  \frac{1}{4} \Th_{mab} \ve^\dag ( 2 \eta^{cb} \g^a - 2 \eta^{ca} \g^b ) \ve\, E_c^n + v^c \del_mE_c^n\\
&\= \left( - \Th_{m}{}^c{}_b  E_c^n e^b_p  +  e_p^b \del_m E_c^n \right)v^p \\
&\= -\G_m{}^n{}_p v^p~,
\end{split}
\eeq
where we have used the usual relations
$$
\{\g^a, \g^b\} \= 2g^{ab}~, \qquad [\g^{ab},\g^c ] \=   2  \eta^{nb}  \g^a - 2 \eta^{na} \g^b ~,
$$
as well as the relation between connection coefficients and the spin connection
\beq
\G_m{}^n{}_p \=   \Th_{m}{}^c{}_b  E_c^n e^b_p  -  e_p^b \del_m E_c^n~.
\eeq
Thus
$$
\del_m v^n + \G_m{}^n{}_p  v^p   \= 0~.
$$

If we return to the gravitino equation and choose $\nabla^+$ so that $\o = \o^\LC + \half H$ then 
$$
\Th^+_{mab} \= \Th^\LC_{mab}+ \half H_{mab}~, \quad \G^+_m{}^n{}_p  \= \G^\LC_m{}^n{}_p + \half H_m{}^n{}_p~,
$$
and
\beq\label{eq:covDeriv}
\nabla_m^\LC v^n+ \half H_m{}^n{}_{p}  v^p \= 0~.
\eeq

\newpage
\subsection{Bergshoeff de-Roo}
In Appendix A of BdR \cite{Bergshoeff:1989de} the action and supersymmetry variations correct to $\ap^2$ are written out. Here we provide a simple dictionary of notation. The left-hand side is always BdR; the right hand side is our notation, largely the same as AQS \cite{Anguelova:2010ed}: 
\beq
\begin{split}
& \phi \= e^{2\Phi / 3}~, \qquad \phi^{-1} \del_m \phi \= \frac{2}{3} \del_m \Phi~, \quad \phi^{-3} \= e^{-2\Phi}~,\\
 &\omega_{mab} \= - \Th^\LC_{mab}~, \quad R(\omega) \= - R(\Th^\LC)~, \\
 &H^{BdR}_{mnp} \= \frac{1}{3\sqrt{2}} H_{mnp}~, \quad B^{BdR}_{mn} \= \frac{1}{\sqrt{2}} B_{mn}~,\\
& \Omega^\pm_{mab} \=\!- \Th^\mp_{mab}~, \qquad \alpha \= -\frac{\ap}{4}~, \quad \beta \= \frac{\ap}{4}~,\\
& T_{mnpq} \= - \frac{1}{3!} (\dd H)_{mnpq}~,\\
& e \= \frac{1}{\k_{10}^2}~,
\end{split}
\eeq
where $\k_{10}$ Newton's constant. 
The Bergshoeff de Roo action becomes
\beq
\begin{split}
 \cL &\= e\phi^{-3} \left\{ -\half R(\o) - \frac{3}{4} H_{mnp} H^{mnp} + \frac{9}{2} (\phi^{-1} \del_m \phi)^3 + \cdots\right\}\\
&\= \frac{e^{-2\Phi}}{2\k_{10}^2  } \left\{ R(\Th^\LC) - \half |H|^2 + 4 (\del_m\Phi)^2 + \cdots\right\} ~.
\end{split}
\eeq
The three-form becomes
\beq
\begin{split}
 H_{mnp} \= 3 \del_{[m} B_{np]}  - \frac{\ap}{4} \CS(A)_{mnp} +  \frac{\ap}{4} \CS(\Th^+)_{mnp}~,
\end{split}
\eeq
while the gravitino and dilatino variations become (correct to $\ap^2$)
\beq
\begin{split}
 \d \Psi_m &\= \left(\del_m + \frac{1}{4} \G^{ab}\left( \Th^-_{mab} + \ap P_{mab} \right)\right)~,\quad P_{mab}\= \frac{1}{4} e^{2\Phi} \nabla^{-\,p} (e^{-2\Phi} (\dd H)_{pmab} \G^{ab} )\ve~,\\
 \d \l &\= -\frac{1}{2\sqrt{2}} \left\{ \delslash \Phi - \half \Hslash + \frac{3\ap}{2}  \Pslash \right\}~, \\
 \d \chi &\=\! -\frac{1}{4} \G^{mn} F_{mn} ~.
 \end{split}
\eeq
As written here there is one important difference as compared with AQS:   in $P_{mab}$ BdR write  the adjoint of $\nabla^-$  contracted on the first index, not the second index as written in AQS. The normalisation of this term differs slightly, $1/4$ here vs $6$ in AQS.

\newpage

\bibliographystyle{utphys}

\begin{thebibliography}{10}

\bibitem{Gunaydin:1995ku}
M.~Gunaydin and H.~Nicolai, ``{Seven-dimensional octonionic Yang-Mills
  instanton and its extension to an heterotic string soliton},''
  \href{http://dx.doi.org/10.1016/0370-2693(95)00375-U}{{\em Phys. Lett. B}
  {\bfseries 351} (1995) 169--172},
  \href{http://arxiv.org/abs/hep-th/9502009}{{\ttfamily arXiv:hep-th/9502009}}.
  [Addendum: Phys.Lett.B 376, 329 (1996)].

\bibitem{Gauntlett:2001ur}
J.~P. Gauntlett, N.~Kim, D.~Martelli, and D.~Waldram, ``{Five-branes wrapped on
  SLAG three cycles and related geometry},''
  \href{http://dx.doi.org/10.1088/1126-6708/2001/11/018}{{\em JHEP} {\bfseries
  11} (2001) 018}, \href{http://arxiv.org/abs/hep-th/0110034}{{\ttfamily
  arXiv:hep-th/0110034}}.

\bibitem{Friedrich:2001nh}
T.~Friedrich and S.~Ivanov, ``{Parallel spinors and connections with skew
  symmetric torsion in string theory},'' {\em Asian J. Math.} {\bfseries 6}
  (2002) 303--336, \href{http://arxiv.org/abs/math/0102142}{{\ttfamily
  arXiv:math/0102142}}.

\bibitem{Friedrich:2001yp}
T.~Friedrich and S.~Ivanov, ``{Killing spinor equations in dimension 7 and
  geometry of integrable G(2) manifolds},''
  \href{http://dx.doi.org/10.1016/S0393-0440(03)00005-6}{{\em J. Geom. Phys.}
  {\bfseries 48} (2003) 1}, \href{http://arxiv.org/abs/math/0112201}{{\ttfamily
  arXiv:math/0112201}}.

\bibitem{Gauntlett:2002sc}
J.~P. Gauntlett, D.~Martelli, S.~Pakis, and D.~Waldram, ``{G structures and
  wrapped NS5-branes},''
  \href{http://dx.doi.org/10.1007/s00220-004-1066-y}{{\em Commun. Math. Phys.}
  {\bfseries 247} (2004) 421--445},
  \href{http://arxiv.org/abs/hep-th/0205050}{{\ttfamily arXiv:hep-th/0205050}}.

\bibitem{Gauntlett:2003cy}
J.~P. Gauntlett, D.~Martelli, and D.~Waldram, ``{Superstrings with intrinsic
  torsion},'' \href{http://dx.doi.org/10.1103/PhysRevD.69.086002}{{\em Phys.
  Rev. D} {\bfseries 69} (2004) 086002},
  \href{http://arxiv.org/abs/hep-th/0302158}{{\ttfamily arXiv:hep-th/0302158}}.

\bibitem{Ivanov:2003nd}
P.~Ivanov and S.~Ivanov, ``{SU(3) instantons and G(2), spin(7) heterotic string
  solitons},'' \href{http://dx.doi.org/10.1007/s00220-005-1396-4}{{\em Commun.
  Math. Phys.} {\bfseries 259} (2005) 79--102},
  \href{http://arxiv.org/abs/math/0312094}{{\ttfamily arXiv:math/0312094}}.

\bibitem{Ivanov:2009rh}
S.~Ivanov, ``{Heterotic supersymmetry, anomaly cancellation and equations of
  motion},'' \href{http://dx.doi.org/10.1016/j.physletb.2010.01.050}{{\em
  Phys.Lett.} {\bfseries B685} (2010) 190--196},
\href{http://arxiv.org/abs/0908.2927}{{\ttfamily arXiv:0908.2927 [hep-th]}}.

\bibitem{Kunitomo:2009mx}
H.~Kunitomo and M.~Ohta, ``{Supersymmetric AdS(3) solutions in Heterotic
  Supergravity},'' \href{http://dx.doi.org/10.1143/PTP.122.631}{{\em Prog.
  Theor. Phys.} {\bfseries 122} (2009) 631--657},
  \href{http://arxiv.org/abs/0902.0655}{{\ttfamily arXiv:0902.0655 [hep-th]}}.

\bibitem{Lukas:2010mf}
A.~Lukas and C.~Matti, ``{G-structures and Domain Walls in Heterotic
  Theories},'' \href{http://dx.doi.org/10.1007/JHEP01(2011)151}{{\em JHEP}
  {\bfseries 01} (2011) 151}, \href{http://arxiv.org/abs/1005.5302}{{\ttfamily
  arXiv:1005.5302 [hep-th]}}.

\bibitem{Gray:2012md}
J.~Gray, M.~Larfors, and D.~Lust, ``{Heterotic domain wall solutions and SU(3)
  structure manifolds},'' \href{http://dx.doi.org/10.1007/JHEP08(2012)099}{{\em
  JHEP} {\bfseries 08} (2012) 099},
  \href{http://arxiv.org/abs/1205.6208}{{\ttfamily arXiv:1205.6208 [hep-th]}}.

\bibitem{ReyesCarrion:1998si}
R.~Reyes~Carrion, ``{A generalization of the notion of instanton},''
  \href{http://dx.doi.org/10.1016/S0926-2245(97)00013-2}{{\em Differ. Geom.
  Appl.} {\bfseries 8} (1998) 1--20}.

\bibitem{fernandez1998dolbeault}
M.~Fern{\'a}ndez and L.~Ugarte, ``Dolbeault cohomology for g2-manifolds,'' {\em
  Geometriae Dedicata} {\bfseries 70} no.~1, (1998) 57--86.

\bibitem{delaOssa:2016ivz}
X.~de~la Ossa, M.~Larfors, and E.~E. Svanes, ``{Infinitesimal moduli of G2
  holonomy manifolds with instanton bundles},''
  \href{http://dx.doi.org/10.1007/JHEP11(2016)016}{{\em JHEP} {\bfseries 11}
  (2016) 016}, \href{http://arxiv.org/abs/1607.03473}{{\ttfamily
  arXiv:1607.03473 [hep-th]}}.

\bibitem{Clarke:2016qtg}
A.~Clarke, M.~Garcia-Fernandez, and C.~Tipler, ``{Moduli of $G_2$ structures
  and the Strominger system in dimension 7},''
  \href{http://arxiv.org/abs/1607.01219}{{\ttfamily arXiv:1607.01219
  [math.DG]}}.

\bibitem{delaOssa:2017pqy}
X.~de~la Ossa, M.~Larfors, and E.~E. Svanes, ``{The Infinitesimal Moduli Space
  of Heterotic G$_{2}$ Systems},''
  \href{http://dx.doi.org/10.1007/s00220-017-3013-8}{{\em Commun. Math. Phys.}
  {\bfseries 360} no.~2, (2018) 727--775},
  \href{http://arxiv.org/abs/1704.08717}{{\ttfamily arXiv:1704.08717
  [hep-th]}}.

\bibitem{delaOssa:2017gjq}
X.~de~la Ossa, M.~Larfors, and E.~E. Svanes, ``{Restrictions of Heterotic $G_2$
  Structures and Instanton Connections},'' in {\em {Nigel Hitchin's 70th
  Birthday Conference}}.
\newblock 9, 2017.
\newblock \href{http://arxiv.org/abs/1709.06974}{{\ttfamily arXiv:1709.06974
  [math.DG]}}.

\bibitem{Fiset:2017auc}
M.-A. Fiset, C.~Quigley, and E.~E. Svanes, ``{Marginal deformations of
  heterotic G$_{2}$ sigma models},''
  \href{http://dx.doi.org/10.1007/JHEP02(2018)052}{{\em JHEP} {\bfseries 02}
  (2018) 052}, \href{http://arxiv.org/abs/1710.06865}{{\ttfamily
  arXiv:1710.06865 [hep-th]}}.

\bibitem{Clarke:2020erl}
A.~Clarke, M.~Garcia-Fernandez, and C.~Tipler, ``{$T$-dual solutions and
  infinitesimal moduli of the $G_2$-Strominger system},''
  \href{http://dx.doi.org/10.4310/ATMP.2022.v26.n6.a3}{{\em Adv. Theor. Math.
  Phys.} {\bfseries 26} no.~6, (2022) 1669--1704},
  \href{http://arxiv.org/abs/2005.09977}{{\ttfamily arXiv:2005.09977
  [math.DG]}}.

\bibitem{Silva:2024fvl}
A.~A.~d. Silva, M.~Garcia-Fernandez, J.~D. Lotay, and H.~N.~S. Earp, ``{Coupled
  $\operatorname{G}_2$-instantons},''
  \href{http://arxiv.org/abs/2404.12937}{{\ttfamily arXiv:2404.12937
  [math.DG]}}.

\bibitem{Anderson:2010mh}
L.~B. Anderson, J.~Gray, A.~Lukas, and B.~Ovrut, ``{Stabilizing the Complex
  Structure in Heterotic Calabi-Yau Vacua},''
  \href{http://dx.doi.org/10.1007/JHEP02(2011)088}{{\em JHEP} {\bfseries 02}
  (2011) 088},
\href{http://arxiv.org/abs/1010.0255}{{\ttfamily arXiv:1010.0255 [hep-th]}}.

\bibitem{Anderson:2011ty}
L.~B. Anderson, J.~Gray, A.~Lukas, and B.~Ovrut, ``{The Atiyah Class and
  Complex Structure Stabilization in Heterotic Calabi-Yau Compactifications},''
  \href{http://dx.doi.org/10.1007/JHEP10(2011)032}{{\em JHEP} {\bfseries 10}
  (2011) 032},
\href{http://arxiv.org/abs/1107.5076}{{\ttfamily arXiv:1107.5076 [hep-th]}}.

\bibitem{Nemeschansky:1986yx}
D.~Nemeschansky and A.~Sen, ``{Conformal Invariance of Supersymmetric Sigma
  Models on Calabi-Yau Manifolds},''
\href{http://dx.doi.org/10.1016/0370-2693(86)91394-8}{{\em Phys.Lett.}
  {\bfseries B178} (1986) 365}.

\bibitem{Jardine:2018sft}
I.~T. Jardine and C.~Quigley, ``{Conformal invariance of (0, 2) sigma models on
  Calabi-Yau manifolds},''
  \href{http://dx.doi.org/10.1007/JHEP03(2018)090}{{\em JHEP} {\bfseries 03}
  (2018) 090}, \href{http://arxiv.org/abs/1801.04336}{{\ttfamily
  arXiv:1801.04336 [hep-th]}}.

\bibitem{Witten:1985bz}
E.~Witten, ``{New Issues in Manifolds of SU(3) Holonomy},''
\href{http://dx.doi.org/10.1016/0550-3213(86)90202-6}{{\em Nucl. Phys.}
  {\bfseries B268} (1986) 79}.

\bibitem{Witten:1986kg}
L.~Witten and E.~Witten, ``{Large Radius Expansion of Superstring
  Compactifications},''
\href{http://dx.doi.org/10.1016/0550-3213(87)90249-5}{{\em Nucl. Phys.}
  {\bfseries B281} (1987) 109--126}.

\bibitem{Garcia-Fernandez:2015hja}
M.~Garcia-Fernandez, R.~Rubio, and C.~Tipler, ``{Infinitesimal moduli for the
  Strominger system and Killing spinors in generalized geometry},''
  \href{http://dx.doi.org/10.1007/s00208-016-1463-5}{{\em Math. Ann.}
  {\bfseries 369} no.~1-2, (2017) 539--595},
  \href{http://arxiv.org/abs/1503.07562}{{\ttfamily arXiv:1503.07562
  [math.DG]}}.

\bibitem{Becker:2014rea}
K.~Becker, D.~Robbins, and E.~Witten, ``{The $\alpha'$ Expansion On A Compact
  Manifold Of Exceptional Holonomy},''
  \href{http://dx.doi.org/10.1007/JHEP06(2014)051}{{\em JHEP} {\bfseries 06}
  (2014) 051}, \href{http://arxiv.org/abs/1404.2460}{{\ttfamily arXiv:1404.2460
  [hep-th]}}.

\bibitem{delaOssa:2021qlt}
X.~de~la Ossa and M.~Galdeano, ``{Families of solutions of the heterotic G$_2$
  system},'' \href{http://arxiv.org/abs/2111.13221}{{\ttfamily arXiv:2111.13221
  [hep-th]}}.

\bibitem{Lotay:2021eog}
J.~D. Lotay and H.~N.~S. Earp, ``{The heterotic G2 system on contact Calabi Yau
  7-manifolds},'' \href{http://dx.doi.org/10.1090/btran/129}{{\em Trans. Am.
  Math. Soc. Ser. B} {\bfseries 10} no.~26, (2023) 907--943},
  \href{http://arxiv.org/abs/2101.06767}{{\ttfamily arXiv:2101.06767
  [math.DG]}}.

\bibitem{Galdeano:2024fsc}
M.~Galdeano and L.~Stecker, ``{The heterotic G$_2$ system with reducible
  characteristic holonomy},'' \href{http://arxiv.org/abs/2403.00084}{{\ttfamily
  arXiv:2403.00084 [math.DG]}}.

\bibitem{delaOssa:2014lma}
X.~de~la Ossa, M.~Larfors, and E.~E. Svanes, ``{Exploring $SU(3)$ structure
  moduli spaces with integrable $G_2$ structures},''
  \href{http://dx.doi.org/10.4310/ATMP.2015.v19.n4.a5}{{\em Adv. Theor. Math.
  Phys.} {\bfseries 19} (2015) 837--903},
  \href{http://arxiv.org/abs/1409.7539}{{\ttfamily arXiv:1409.7539 [hep-th]}}.

\bibitem{McOrist:2021dnd}
J.~McOrist and E.~E. Svanes, ``{Heterotic quantum cohomology},''
  \href{http://dx.doi.org/10.1007/JHEP11(2022)096}{{\em JHEP} {\bfseries 11}
  (2022) 096}, \href{http://arxiv.org/abs/2110.06549}{{\ttfamily
  arXiv:2110.06549 [hep-th]}}.

\bibitem{McOrist:2024zdz}
J.~McOrist, S.~Picard, and E.~E. Svanes, ``{A Heterotic Hermitian--Yang--Mills
  Equivalence},'' {\em Accepted Comm. Math. Phys.} (2, 2024) ,
  \href{http://arxiv.org/abs/2402.10354}{{\ttfamily arXiv:2402.10354
  [hep-th]}}.

\bibitem{McOrist:2019mxh}
J.~McOrist and R.~Sisca, ``{Small gauge transformations and universal geometry
  in heterotic theories},''
  \href{http://dx.doi.org/10.3842/SIGMA.2020.126}{{\em SIGMA} {\bfseries 16}
  (2020) 126}, \href{http://arxiv.org/abs/1904.07578}{{\ttfamily
  arXiv:1904.07578 [hep-th]}}.

\bibitem{Melnikov:2011ez}
I.~V. Melnikov and E.~Sharpe, ``{On marginal deformations of (0,2) non-linear
  sigma models},'' \href{http://dx.doi.org/10.1016/j.physletb.2011.10.055}{{\em
  Phys.Lett.} {\bfseries B705} (2011) 529--534},
\href{http://arxiv.org/abs/1110.1886}{{\ttfamily arXiv:1110.1886 [hep-th]}}.

\bibitem{Anderson:2014xha}
L.~B. Anderson, J.~Gray, and E.~Sharpe, ``{Algebroids, Heterotic Moduli Spaces
  and the Strominger System},''
  \href{http://dx.doi.org/10.1007/JHEP07(2014)037}{{\em JHEP} {\bfseries 07}
  (2014) 037},
\href{http://arxiv.org/abs/1402.1532}{{\ttfamily arXiv:1402.1532 [hep-th]}}.

\bibitem{delaOssa:2014msa}
X.~de~la Ossa and E.~E. Svanes, ``{Connections, Field Redefinitions and
  Heterotic Supergravity},''
  \href{http://dx.doi.org/10.1007/JHEP12(2014)008}{{\em JHEP} {\bfseries 12}
  (2014) 008},
\href{http://arxiv.org/abs/1409.3347}{{\ttfamily arXiv:1409.3347 [hep-th]}}.

\bibitem{Eguchi:1980jx}
T.~Eguchi, P.~B. Gilkey, and A.~J. Hanson, ``{Gravitation, Gauge Theories and
  Differential Geometry},''
\href{http://dx.doi.org/10.1016/0370-1573(80)90130-1}{{\em Phys. Rept.}
  {\bfseries 66} (1980) 213}.

\bibitem{Bergshoeff:1989de}
E.~Bergshoeff and M.~de~Roo, ``{The Quartic Effective Action of the Heterotic
  String and Supersymmetry},''
\href{http://dx.doi.org/10.1016/0550-3213(89)90336-2}{{\em Nucl.Phys.}
  {\bfseries B328} (1989) 439}.

\bibitem{Bergshoeff:1988nn}
E.~Bergshoeff and M.~de~Roo, ``{Supersymmetric Chern-Simons Terms in
  Ten-Dimensions},''
\href{http://dx.doi.org/10.1016/0370-2693(89)91420-2}{{\em Phys.Lett.}
  {\bfseries B218} (1989) 210}.

\bibitem{McOrist:2010jw}
J.~McOrist, D.~R. Morrison, and S.~Sethi, ``{Geometries, Non-Geometries, and
  Fluxes},'' \href{http://dx.doi.org/10.4310/ATMP.2010.v14.n5.a4}{{\em Adv.
  Theor. Math. Phys.} {\bfseries 14} no.~5, (2010) 1515--1583},
  \href{http://arxiv.org/abs/1004.5447}{{\ttfamily arXiv:1004.5447 [hep-th]}}.

\bibitem{Strominger:1986uh}
A.~Strominger, ``{Superstrings with Torsion},''
  \href{http://dx.doi.org/10.1016/0550-3213(86)90286-5}{{\em Nucl. Phys. B}
  {\bfseries 274} (1986) 253}.

\bibitem{Hull:1986kz}
C.~Hull, ``{Compactifications of the Heterotic Superstring},''
\href{http://dx.doi.org/10.1016/0370-2693(86)91393-6}{{\em Phys.Lett.}
  {\bfseries B178} (1986) 357}.

\bibitem{Candelas:2016usb}
P.~Candelas, X.~de~la Ossa, and J.~McOrist, ``{A Metric for Heterotic
  Moduli},'' \href{http://dx.doi.org/10.1007/s00220-017-2978-7}{{\em Commun.
  Math. Phys.} {\bfseries 356} no.~2, (2017) 567--612},
\href{http://arxiv.org/abs/1605.05256}{{\ttfamily arXiv:1605.05256 [hep-th]}}.

\bibitem{Atiyah:1955}
M.~F. Atiyah, ``Complex analytic connections in fibre bundles,'' {\em Trans.
  AMS} {\bfseries 85} no.~1, (1957) 181--207.

\bibitem{Chisamanga:2024xbm}
B.~Chisamanga, J.~McOrist, S.~Picard, and E.~E. Svanes, ``{The decoupling of
  moduli about the standard embedding},''
  \href{http://dx.doi.org/10.1007/JHEP01(2025)032}{{\em JHEP} {\bfseries 01}
  (2025) 032}, \href{http://arxiv.org/abs/2409.04350}{{\ttfamily
  arXiv:2409.04350 [hep-th]}}.

\bibitem{deLazari:2024zkg}
H.~de~L\'azari, J.~D. Lotay, H.~S. Earp, and E.~E. Svanes, ``{Local
  descriptions of the heterotic SU(3) moduli space},''
  \href{http://arxiv.org/abs/2409.04382}{{\ttfamily arXiv:2409.04382
  [math.DG]}}.

\bibitem{Ashmore:2018ybe}
A.~Ashmore, X.~De~La~Ossa, R.~Minasian, C.~Strickland-Constable, and E.~E.
  Svanes, ``{Finite deformations from a heterotic superpotential: holomorphic
  Chern-Simons and an $L_\infty$ algebra},''
  \href{http://dx.doi.org/10.1007/JHEP10(2018)179}{{\em JHEP} {\bfseries 10}
  (2018) 179}, \href{http://arxiv.org/abs/1806.08367}{{\ttfamily
  arXiv:1806.08367 [hep-th]}}.

\bibitem{Candelas:2018lib}
P.~Candelas, X.~De~La~Ossa, J.~McOrist, and R.~Sisca, ``{The Universal Geometry
  of Heterotic Vacua},'' \href{http://dx.doi.org/10.1007/JHEP02(2019)038}{{\em
  JHEP} {\bfseries 02} (2019) 038},
  \href{http://arxiv.org/abs/1810.00879}{{\ttfamily arXiv:1810.00879
  [hep-th]}}.

\bibitem{Anguelova:2010ed}
L.~Anguelova, C.~Quigley, and S.~Sethi, ``{The Leading Quantum Corrections to
  Stringy Kahler Potentials},''
  \href{http://dx.doi.org/10.1007/JHEP10(2010)065}{{\em JHEP} {\bfseries 1010}
  (2010) 065},
\href{http://arxiv.org/abs/1007.4793}{{\ttfamily arXiv:1007.4793 [hep-th]}}.
"

\bibitem{Becker:2007zj}
K.~Becker, M.~Becker, and J.~H. Schwarz, {\em {String theory and M-theory: A
  modern introduction}}.
\newblock Cambridge University Press,
2006.
\newblock

\bibitem{delaOssa:2014cia}
X.~de~la Ossa and E.~E. Svanes, ``{Holomorphic Bundles and the Moduli Space of
  N=1 Supersymmetric Heterotic Compactifications},''
  \href{http://dx.doi.org/10.1007/JHEP10(2014)123}{{\em JHEP} {\bfseries 10}
  (2014) 123},
\href{http://arxiv.org/abs/1402.1725}{{\ttfamily arXiv:1402.1725 [hep-th]}}.

\bibitem{Martelli:2010jx}
D.~Martelli and J.~Sparks, ``{Non-Kahler heterotic rotations},''
  \href{http://dx.doi.org/10.4310/ATMP.2011.v15.n1.a4}{{\em
  Adv.Theor.Math.Phys.} {\bfseries 15} (2011) 131--174},
\href{http://arxiv.org/abs/1010.4031}{{\ttfamily arXiv:1010.4031 [hep-th]}}.

\end{thebibliography}
\providecommand{\href}[2]{#2}\begingroup\raggedright\endgroup

\end{document}